\documentclass[journal,draftclsnofoot,onecolumn,12pt]{IEEEtran}  
\usepackage{amsthm,amssymb,graphicx,multirow,amsmath,color,amsfonts}

\usepackage[update,prepend]{epstopdf}
\usepackage[noadjust]{cite}
\usepackage[latin1]{inputenc}
\usepackage{tikz}
\usetikzlibrary{angles, quotes}
\usepackage{bbm} 
\usepackage{pdfpages}
\usepackage{tabulary}
\usepackage{multirow}
\usepackage{comment}
\usepackage{dsfont}
\usetikzlibrary{calc}
\usepackage{ amssymb }
\usepackage[mathscr]{eucal}
\usepackage{bm}
\usepackage{diagbox}

\newtheorem{corollary}{Corollary}

\newcommand{\integral}{\int\limits}

\newcommand{\ind}{\mathds{1}}
\newcommand{\blank}{\!\;}
\renewcommand{\Bar}{\bm\bar}
\renewcommand{\Check}{\bm\check}
\renewcommand{\P}{\mathbb{P}}
\renewcommand{\it}{\textit}

\def\nb0{{\mathbf{0}}}
\def\nb1{{\mathbf{1}}}

\newtheorem{lemma}{Lemma}

\newtheorem{theorem}{Theorem}
\newtheorem{prop}{Proposition}

\def\argmax{\operatorname{arg~max}}
\def\E{\mathbb{E}}

\def\P{\mathbb{P}}

\def\R{\mathbb{R}}

\allowdisplaybreaks 

\begin{document}
\graphicspath{{./Figures/}}
\title{UAV-Aided Post-Disaster Cellular Networks:\\A Novel Stochastic Geometry Approach}
\author{Maurilio Matracia, \normalfont Mustafa A. Kishk,  \normalfont and Mohamed-Slim Alouini
\thanks{
Maurilio Matracia and Mohamed-Slim Alouini are with the Computer, Electrical, and Mathematical Sciences and Engineering (CEMSE) Division at King Abdullah University of Science and Technology (KAUST), Thuwal, Kingdom of Saudi Arabia (KSA) (email: $\{$maurilio.matracia; slim.alouini$\}$@kaust.edu.sa).\par
Mustafa Kishk is with the Department of Electronic Engineering, National University of Ireland, Maynooth, W23 F2H6, Ireland (email: mustafa.kishk@mu.ie).
}}

\maketitle
\vspace{-3mm}

\begin{abstract}
Motivated by the need for ubiquitous and reliable communications in post-disaster emergency management systems (EMSs), we hereby present a novel and efficient stochastic geometry (SG) framework.
This mathematical model is specifically designed to evaluate the quality of service (QoS) experienced by a typical ground user equipment (UE) residing either inside or outside a generic area affected by a calamity.
In particular, we model the functioning terrestrial base stations (TBSs) as an inhomogeneous Poisson point process (IPPP), and assume that a given number of uniformly distributed unmanned aerial vehicles (UAVs) equipped with cellular transceivers is deployed in order to compensate for the damage suffered by some of the existing TBSs.
The downlink (DL) coverage probability is then derived based on the maximum average received power association policy and the assumption of Nakagami-\it{m} fading conditions for all wireless links.
The proposed numerical results show insightful trends in terms of coverage probability, depending on: distance of the UE from the disaster epicenter, disaster radius, quality of resilience (QoR) of the terrestrial network, and fleet of deployed ad-hoc aerial base stations (ABSs).
The aim of this paper is therefore to prove the effectiveness of vertical heterogeneous networks (VHetNets) in emergency scenarios, which can both stimulate the involved authorities for their implementation and inspire researchers to further investigate related problems. 
\end{abstract}

\vspace{0mm}


\section{Introduction} \label{sec:Intro}
Disasters represent of the main threats to modern communities, because they can potentially compromise every form of life within the region involved, apart from the risk of damaging its economy and cultural heritage.
Contextually, the United Nations (UN) established the International Search and Rescue Advisory Group (INSARAG) in 1991 in order to essentially~\cite{INSARAG}:\\
\textit{(i)} Improve the effectiveness of emergency preparedness and response operations;\\
\textit{(ii)} Design activities that improve search-and-rescue (SAR) missions in disaster-prone countries;\\
\textit{(iii)} Ameliorate cooperation among international urban-SAR (USAR) teams and develop procedures and systems for national teams operating internationally;\\
\textit{(iv)} Develop USAR procedures, guidelines and best practices for the emergency relief phase. \par
Emergency situations often require reliable cellular coverage over large areas to ensure the safety of victims and first responders (FRs), especially during SAR missions.
However, telecom infrastructure dysfunction (\it{e.g.}, failure or lack of power supply) is one of the main concerns related to current network architectures.
Indeed, the quality of telecommunications usually decreases after the occurrence of a disaster.
Perturbations to the networking equipment can often lead to continuous reconfiguration of the routing tables, a larger ratio of packet losses, disturbances to radio frequency (RF) signals, and many other issues \cite{esposito20reinforced, matracia22survey}.
Consequently, ABSs consisting of UAVs equipped with cellular transceivers are gaining more and more attention as an alternative solution for supporting TBSs in post-disaster scenarios \cite{matracia22survey, Matracia21disaster}. 
The latter, in fact, are generally susceptible to earthquakes, tornadoes, explosions, and many other serious perturbations. \par
Apart from their mobility, using ABSs as ad-hoc nodes in emergency situations is also more appropriate than using cell towers because of their lower cost, faster deployment, and higher altitude \cite{Mozaffari19tutorial}.
By reaching a higher altitude, indeed, it is possible for a BS to achieve a larger footprint as well as a higher probability of establishing line-of-sight (LoS) transmissions, which can generally lead to better communication channels compared to the case of non-LoS (NLoS) transmissions \cite{al2014optimal}.
Furthermore, promising advancements in avionics and especially drone technology have enabled the use of such vehicles for several purposes (including disaster monitoring \cite{savkin20navigation}, damage assessment \cite{wu18coupling}, and first aid and supply delivery \cite{sanjana20aid}, for instance), although their flight time is considerably reduced whenever operating in multi-task mode.
However, there are many types of vehicles that can be used as ABSs: these are usually categorized as low-altitude platforms (LAPs) and high-altitude platforms (HAPs) \cite{Matracia21disaster}.
Drones (either tethered \cite{Kishk20magazine, kishk20placement} or untethered) and tethered balloons are common examples of LAPs, and usually their altitude does not exceed $10\,$km.
On the other hand, airships, gliders, and untethered balloons fall in the category of HAPs, since they are usually designed to operate in the stratosphere. \par
In this paper, we consider both LAPs and HAPs as a potential solution for supporting post-disaster communications while capturing the resiliency of the terrestrial cellular infrastructure.
{ Given the inherent randomness of the network nodes' deployment and resilience, for our analysis we decided to implement an SG approach due to its tractability and accuracy.} \par
More details on the contributions of this work are provided in Sec.~\ref{subsec:contributions}.

\subsection{Related Works}
This subsection provides a concise summary of the relevant literature works on UAV-assisted disaster communications and SG-based analysis of UAV networks. 
\subsubsection{UAV-Aided Disaster Communications} 
As explained in~\cite{Erdelj16_disasterManagement}, the importance of UAVs in emergency scenarios is not limited to post-disaster situations but also concerns the phases of pre-disaster preparedness and disaster assessment.
Indeed, many works in the literature discussing disaster communications have considered using UAVs for applications related to situational awareness \cite{Graven17managing}, damage assessment \cite{ezequiel2014uav}, and network rehabilitation \cite{Naqvi18key,Selim18battery,Matracia21disaster}.
Authors in~\cite{Graven17managing}, in fact, approached the problem of situational awareness by deploying drones in order to capture a digital terrain model and place sensors in a disaster-struck area, creating a dynamic sensor network.
On the other hand, \cite{ezequiel2014uav} proposed combining UAV-based imagery with ground observations and collaborative sharing with domain experts for either post-disaster assessment, environmental management, or monitoring of infrastructure development.
However, the most interesting application for UAVs in disaster scenarios is probably to support or even substitute the terrestrial cellular infrastructure, as respectively suggested in~\cite{Naqvi18key} and \cite{Selim18battery}.\par
{ Finally, it is worth mentioning the current research interest in achieving UAVs' minimum energy consumption and optimal placement.
For example, the work presented in~\cite{shakoor19role} introduced the first multi-layered heterogeneous network architecture that integrates ad hoc UAVs into public safety communications; in particular, said architecture is expected to enable reliable communications in basements by means of both wired and wireless links.
On the other side, in~\cite{masroor21efficient} a novel multi-objective integer linear optimization problem (ILP) was solved in order to optimally deploy the UAVs assisting disaster-affected users;
the authors compared the branch-and-bound (B$\&$B) algorithm with their proposed low-complexity heuristic one.}\par
For a more detailed overview of this topic, the reader can refer to Ref.~\cite{matracia22survey}.

\subsubsection{SG for UAV-Assisted Networks} 
During the last decade, SG has emerged in the literature as one of the most effective mathematical tools for modeling and analyzing large scale VHetNets.
More specifically, the performances of UAV-assisted terrestrial cellular networks have been evaluated via SG approaches in works such as \cite{arshad2018integrating,Alzenad19,Matracia21rural,Kouzayha20hybrid,hayajneh18performance}. \par
Arshad \it{et al.}~\cite{arshad2018integrating} proposed an architecture consisting of macro and small TBSs supported by ABSs for evaluating the QoS experienced by either stationary or mobile users (by taking into account the effect of handover rates).
Moreover, a setup with TBSs and ABSs modeled by means of distinct homogeneous Poisson point processes (HPPPs) was introduced in~\cite{Alzenad19} in order to derive both the coverage probability and average data rate experienced by a typical ground UE.
Following the same lines, in~\cite{Matracia21rural} we used two different Poisson point processes (PPPs) to model the aerial and terrestrial nodes, and introduced specific features such as the aerial exclusion zone and the inhomogeneous distribution of the TBSs' density to accurately model comprehensive environments that include both urban and exurban areas. \par
Furthermore, authors in~\cite{Kouzayha20hybrid} relied on SG to evaluate the effectiveness of ABSs, modeled as a binomial point process (BPP), while taking into account also the backhaul probability.
Finally, we consider \cite{hayajneh18performance} as the most related work since it is the only one modeling also the resilience of the terrestrial nodes, which is done by introducing a thinning probability for the PPP-distributed TBSs.
However, for the sake of simplicity, the latter work assumed the damages to spread over the entire ground plane, which may not be accurate for typical post-disaster scenarios.

\subsection{Contributions} \label{subsec:contributions}
The contributions of our paper involve multiple aspects, as explained in this subsection. \par
\subsubsection{System Model}
We consider a large-scale post-disaster wireless network consisting of both terrestrial and aerial nodes.
We devise accurate inhomogeneous PPPs (IPPPs) to model the planar distribution of the functioning TBSs { (that is, we assume that the original distribution is thinned according to a certain probability depending on the distance from the disaster epicenter)}, both inside and outside a circular disaster-struck zone, whereas the aerial network is modeled as a BPP confined to the vertical projection of the disaster area. \par
Thus, the length of the disaster radius and the behavior of the QoR of the terrestrial network represent crucial parameters since they allow to capture the severity of any catastrophic event~\cite{Matracia21disaster}. 
This, in turn, has an influence on the optimal fleet of ABSs (identified by number and the type of ad hoc nodes required to provide the highest QoS).\par
{ In conclusion, we consider our system model as a contribution to the existing literature because it takes into account both the vertical heterogeneity (due to the presence of aerial and terrestrial
BSs) and the horizontal heterogeneity (due to the distribution of the surviving TBSs and the consequent placement of the ABSs) of integrated post-disaster wireless networks in an original way.}

\subsubsection{Performance Analysis}
To the best of our knowledge, this paper provides the first SG-based framework specifically designed to analyze the DL performances of 5G and beyond cellular VHetNets affected by a localized disaster while taking into account the QoR of the terrestrial infrastructure. Also, our framework is more general compared to the baseline ones \cite{afshang17bpp,Chetlur3D_bpp}, since it allows to evaluate the performance of the network even when the user is outside the ground projection of the considered BPP's domain.
\par
More specifically, the devised framework introduces a new method that makes use of indicator functions in order to avoid bulky piecewise expressions for describing the novel cumulative distribution functions (CDFs), probability density functions (PDFs), and Laplace transforms of the interference derived for each layer.
In other words, compared to the methods available in the literature \cite{Alzenad19,Chetlur3D_bpp,Matracia21rural}, this one better conveys the meaning of the derived expressions and eases their numerical implementation. \par
We applied our method to compute the \it{spatial} coverage probability (that is, we focus on covering an area independently from the actual distribution of the users), and validated the results via Monte Carlo simulations. 
In addition to better conveying the meaning of the derived expressions and easing their numerical implementation, another advantage of our method is its generality: indeed, the proposed expressions hold irrespective of the UE's location, whereas the conventional approaches would require different expressions depending on whether the typical user resides inside or outside the disaster-struck area.

\subsubsection{System-Level Insights}
Several fruitful insights can be extracted by investigating the behavior of the coverage probability in response to the considered parameters.
For example, the obtained results show that the type and cardinality of a fleet of ABSs have a strong influence on the coverage probability, and should be optimized based on topological aspects such as the state of the terrestrial infrastructure, the disaster radius, and the typical UE's location. 
Indeed, even when neglecting the strict technological and economic constraints (\it{e.g.}, UAVs' autonomy and backhaul, as well as their availability and associated cost of deployment), exploiting dense VHetNets imposes a trade-off between offering a strong desired signal and causing considerable interference to the UE.

\section{System Model} \label{sec:SystemModel}
\subsection{Network Model}
We consider a post-disaster scenario where the DL cellular network infrastructure is affected by a disaster, and thus a fleet of ad-hoc ABSs is deployed in order to make up for the failure of some TBSs within the suffered region.
For the sake of both conciseness and readability we introduce a specific notation for the types of BSs, in accordance with Table~\ref{tab:subscripts} (where $Q$ denotes the average received power and $W$ the location of the BS).

\begin{table*} 
  \begin{center}
    \caption{Main Subscripts}
    \vspace{-3mm}
    \label{tab:subscripts}
     \begin{tabular}{|c|c|c|} \hline
      \textbf{Notation} & \textbf{Description} & \textbf{Definition}  \\
      \hline
      $A$ & ABSs & \textbf{---} \\ \hline 
      $T$ &              Functioning          TBSs & \textbf{---} \\ \hline
      $O$ & Generic type of BSs & $A\oplus T$ \\ \hline
      $B$ & Type of tagged BS & $B\!\in\!\{A,T\} \;\land\; Q_B^*\!>\! Q_O^*\blank,\,\, \forall O\!\neq\! B$ \\ \hline
      $C$ & Type of interfering BSs & $C\!\in\!\{A,T\}\;\land\; Q_{C,W_i}\!<\!Q_B^*\blank,\,\,\forall W_i\!\neq\! W^*$ \\ \hline
      
    \end{tabular}  
  \end{center} 
\end{table*} 

Without any loss of generality, we set the origin $\boldsymbol{O}$ at the epicenter of the disaster.
The disaster area is assumed with altitude $0$, circular with radius $r_d$, and can thus be expressed as $\mathscr{A}_d=\textbf{b}(\boldsymbol{O},0,r_d)\subset\R_0^2\blank$, where $\R_0^2$ is the Euclidean ground plane. \par
{ As in the absence of any calamity the TBSs' planar distribution can generally be modeled by means of an HPPP~\cite{Andrews11,Alzenad19} of intensity $\lambda_0>0$, we hereby assume that the original infrastructure experience random failures within the disaster-struck area.
Therefore, the IPPP $\Phi_{T}\equiv\{Y_i\} \!\subseteq\! \R_0^2\blank$ describes the surviving TBSs' distribution;}
the intensity of this process is  $\lambda_T(r)\!=\!\lambda_0\,\big(\chi(r)\,\ind(r\leq r_d)\!+\!\ind(r>r_d)\big)$, where $r>0$ represents the horizontal distance from the origin and $\chi(r)\in[0,1]$ identifies the QoR of the terrestrial network.\par
\begin{figure*}
\centering
\includegraphics[width=0.95\textwidth, trim={0cm 0.8cm 0cm 1.5cm},clip]{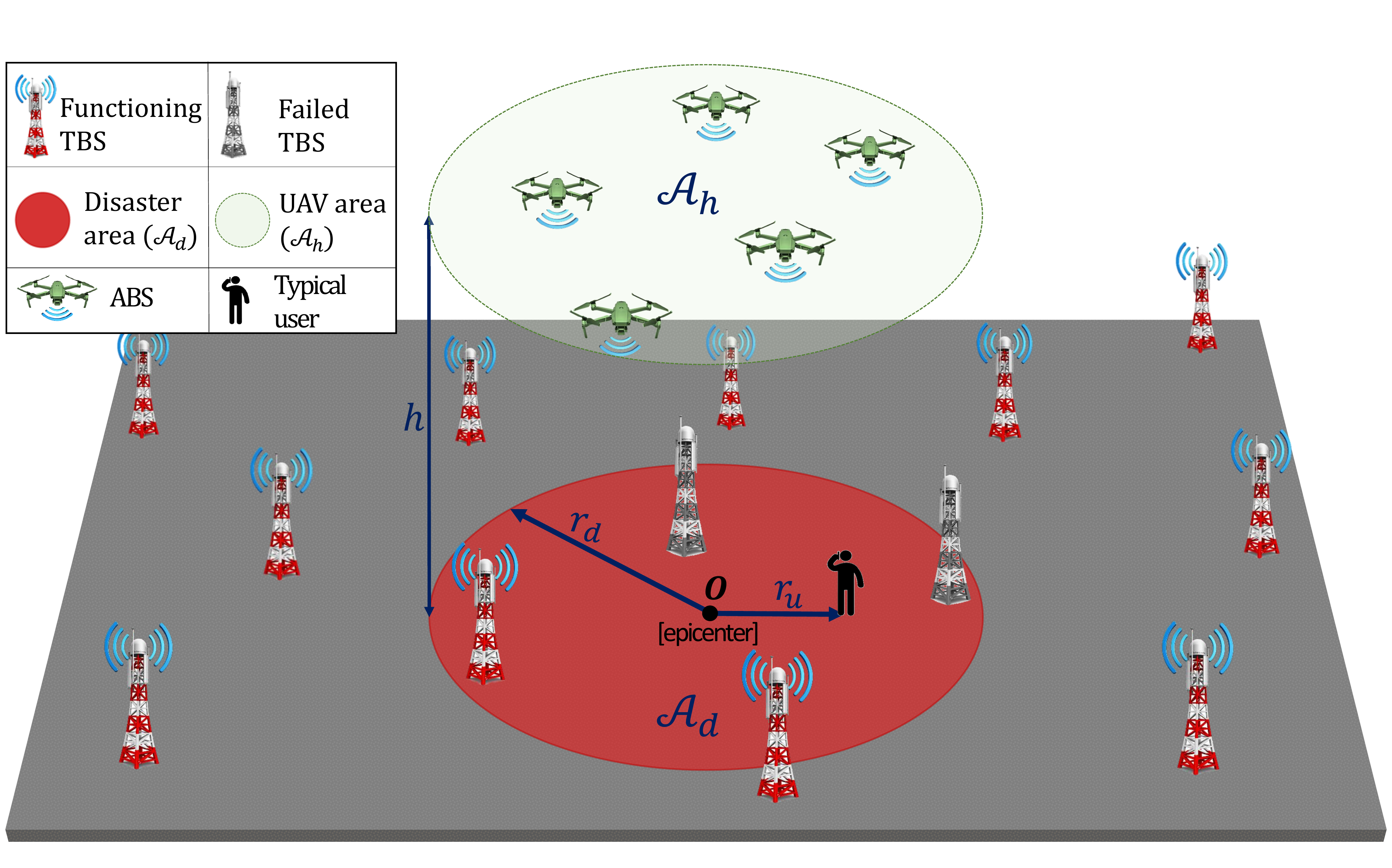}
\vspace{-5mm}
\caption{Schematic representation of the system setup considered: the typical user is located at distance $r_u$ from the epicenter of the disaster (\it{i.e.}, the origin) and associates to the BS that provides the maximum average received power. 
All the failed TBSs belong to a circular disaster-struck region $\mathscr{A}_d$, whereas a fixed number of UAVs reside within its projection $\mathscr{A}_h$.}
\label{fig:system}
\end{figure*}

 
{ Finally, since the number of deployed UAVs decided by the authority is supposedly deterministic},
the ABSs' planar distribution is described by means of a uniform binomial point process (BPP) $\Phi_A\equiv\{X_i\}\!\subseteq\!\mathscr{A}_h$, where $\mathscr{A}_h=\textbf{b}(\boldsymbol{O}, h, r_d)$ indicates the vertical projection at altitude $h$ of $\mathscr{A}_d$ (see Fig.~\ref{fig:system}). 
{ Although ABSs may definitely be subject to failures (especially in case of harsh weather conditions), we henceforth consider them totally resilient, because their deployment would occur after the actual disaster.}

\subsection{Channel model}
This subsection aims to characterize both the terrestrial and aerial wireless channels.
Keeping in mind Table~\ref{tab:subscripts}, we assume that the signals transmitted by any BSs belonging to a given tier $O$ have a fixed, constant transmit power $\rho_O$ and experience standard power-law path loss propagation with path loss exponent ${\alpha_O}\!\geq\!2\!\;$. \par
Let $\eta$ denote the mean additional transmission losses, then we can define $\xi_O=\eta_O\,\rho_O\blank$.
We assume both the terrestrial and aerial links experience small-scale fading in the form of a Nakagami-\it{m} distribution with generic shape parameter $m_O\blank$. 
Note that small-scale fadings are usually Rayleigh or Rician distributed.
However, the Nakagami-\it{m} distribution with shape parameter $m=\frac{(K+1)^2}{2K+1}$ (and scale parameter equal to its reciprocal) allows a fair approximation of the Rician distribution with factor $K$ \cite{Alzenad19}.
For every $W_i\in \Phi_O\blank$, the channel fading power gains $G_{O,W_i}$'s follow a Gamma distribution with PDF given by
\begin{align}
f_{G_O,W_i}(g) = \frac{m_O^{m_O}\,g^{m_O-1}}{\Gamma(m_O)}\,e^{-m_O\,g},    
\end{align}
where $\Gamma(m)=\integral_0^\infty x^{m-1}\,e^{-x}\,{\rm d}x$ identifies the Gamma function. \par
For a given tier $O$, let $Q_O$ denote the random variable referring to the average power received by the typical UE.
We define $Q_O^*$ and $Q_{O,W_i}$ as the received powers coming from the closest and any generic $O$-BSs located at point $W_i\blank$, respectively.
Thus, the random power received by the typical user from a BS located at $W_i$ can be expressed as
\begin{align}
Q_{O,W_i}=\xi_O\,G_{O,W_i}\,(1+D_{W_i})^{-\alpha_O}\approx\xi_O\,G_{O,W_i}\,D_{W_i}^{-\alpha_O},    
\end{align}
where we introduced the \it{modified} path loss to formally avoid the absurdity $Q_{O,W_i}>\xi_O\blank$, occurring for $D_{W_i}<1\,$m; nonetheless, we will (fairly) use the above approximation as we are considering large scale networks.
Finally note that if $O=T$ then $D_{W_i}=\Omega_{W_i}\blank$.

\subsection{Association Policy}
In this paper, the strongest average received power association rule is adopted, meaning that the user connects to the BS providing the highest average received power.
This, however, does not exclude the possibility of having interferers providing higher received powers during a given instant.
Moreover, due to the fact that each type of BS is characterized by a specific path-loss exponent, mean additional transmit losses, and transmit power, the serving BS is guaranteed to be the closest BS but only among the BSs of the same type. \par
Finally, we assume the expected values of the fading gains over all the sets of BSs (\it{i.e.}, $\mathbb{E}[G_{O,W_i}], \forall W_i\in\Phi_O$) to equal $1$.
Hence, the location of the tagged BS will be simply provided by the maximum product $\xi_O\,D_{O,W_i}^{\alpha_O}\blank$, that is \begin{align}
W^*=\underset{{W_i\in\Phi_O}}{\argmax} \,(\xi_O\,D_{O,W_i}^{-\alpha_O})\blank,\,\forall\, O\in\{A,T\}\blank.
\end{align}

\subsection{Interference and Signal-to-Interference-plus-Noise Ratio ($\rm SINR$)} 
The instantaneous $\rm SINR$ can be expressed as 
\begin{align}
    {\rm SINR}=\frac{Q_B^*}{\sigma_n^2+I}\,, 
\end{align}
where $\sigma_n^2$ is the additive white Gaussian noise (AWGN) power and $I$ is the aggregate interference power.
Letting $C$ denote the layer hosting each interfering BS and assuming that all BSs share the same frequency or time resource blocks, then the random variable (RV) $I$ can be introduced as
\begin{align}
    I=\sum\limits_{C=\{A,T\}}\,\sum\limits_{\substack{W_i\in\Phi_C \\ {W_i\neq W^*}}} Q_{C,W_i}\blank.
\end{align}

\subsection{Coverage Probability}
The coverage probability is defined as the complementary cumulative distribution function (CCDF) of the $\rm SINR$ evaluated at a designated threshold $\tau$ ensuring reliable decoding, that is
\begin{align}
    P_c=\P(\rm{SINR}>\tau)\blank.
\end{align}

\section{Performance Analysis} \label{sec:PerformanceAnalysis}
In this section, the distributions of the distance to the closest $O$-BS, the association probabilities, and the conditional Laplace transforms of the interference will be derived for both the aerial and terrestrial layers of BSs in order to obtain the approximate and exact expressions of the coverage probability.

\subsection{Distance to the Nearest $O$-BS}
Intuitively, the coverage probability is a function of the distance between the UE and the tagged BS.
In order to derive the exact and approximate expressions of the coverage probability, the theorems in this subsection characterize the distribution of the horizontal distance between the UE and each closest $O$-BS by computing its CDF.
Consequently, the respective PDF will be derived in a corollary.

\begin{theorem}  \label{thm:CDF_T}
Let $r_u$ be the distance between the typical user and the center of a disaster with radius $r_d\blank$, then the CDF of the random horizontal distance\footnote{$\,$Whenever not specified, we always refer to the distance from the typical UE, around which the polar coordinate system $(\omega,\beta)$ is centered.} 
$Z_T$ between the UE and the closest TBS in an IPPP with density $\lambda_T(r)$ is given by
\begin{align}
    F_{Z_T}(z) =& \, 1 - \exp\bigg( -\integral_0^{2\pi} \integral_0^z \lambda_T\big(r_\Omega(\omega,\beta)\big)\,\omega\,{\rm d}\omega\,{\rm d}\beta\bigg),
\end{align}
where $r_\Omega(\omega,\beta)=\sqrt{r_u^2+\omega^2-2\,r_u\,\omega\,\cos\beta}$ describes the ground distance from the origin. 
\end{theorem}

\begin{IEEEproof}See Appendix~\ref{appx:CDF_T}. \end{IEEEproof}

\begin{corollary}\label{cor:PDF_T}
Henceforth, let the overline characterize the complementary functions (\it{i.e.}, $\Bar{F}_{Z_O}(z)=1-F_{Z_O}(z)$).
The PDF of the distance $Z_T$ between the UE and the closest surviving TBS is 
\begin{align}
    f_{Z_T}(z) = z\,\Bar{F}_{Z_T}(z) \integral_0^{2\pi} \lambda_T\big(r_\Omega(z,\beta)\big)\,\rm{d}\beta\blank.
\end{align}
\end{corollary}

\begin{IEEEproof} See Appendix~\ref{appx:PDF_T}. \end{IEEEproof}

\vspace{4mm}
Due to the inherent complexity of the BPP, the distance distribution to the closest ABS cannot be computed directly.
Therefore, as an intermediate step, we now leverage a well-known property of BPPs in order to obtain the distribution of the horizontal distance $\Omega_A$ between the UE and any ABS.

\begin{prop}
For a given set of $N$ points uniformly distributed over an area $\mathscr{A}$, the points residing in any subarea $\Sigma\subseteq\mathscr{A}$ are uniformly distributed with cardinality $n\sim {\rm Bin}\big(N,\frac{\Sigma}{\mathscr{A}}\big)\,$ \cite[Theorem 2.9]{haenggi12stochastic}.
\label{prop:Bin}
\end{prop}

\begin{lemma}
In accordance to \cite[Lemma 1]{Chetlur3D_bpp}, the horizontal distances $\Omega_A$'s to the set of independently and uniformly distributed UAVs are independent and identically distributed (iid), with the CDF and PDF of each element respectively given by 
\begin{align}
F_{\Omega_A}(\omega) = \frac{\Sigma(\omega)}{\mathscr{A}_d}\,,
\end{align}
and
\begin{align}
    f_{\Omega_A}(\omega) = \frac{1}{\mathscr{A}_d}\,\frac{{\rm d}\Sigma(\omega)}{{\rm d}\omega}\,,
\end{align}
in which 
$\Sigma(\omega)=\integral_0^{2\pi} \integral_0^\omega \ind\big(r_\Omega(\omega',\beta)\!<\!r_d\big)\,\omega'\,{\rm d}\omega'\,{\rm d}\beta$ describes the intersection area between $\mathscr{A}_d$ and the disc of radius $\omega$ centered around the UE, 
and
$\frac{{\rm d}\Sigma(\omega)}{{\rm d}\omega} = \omega \integral_0^{2\pi} \ind\big(r_\Omega(\omega,\beta)\!<\!r_d\big)\,{\rm d}\beta\blank$.
\end{lemma}
\begin{IEEEproof} The expression of $\frac{{\rm d}\Sigma(\omega)}{{\rm d}\omega}$ can be easily derived by applying the Leibniz rule to $\Sigma(\omega)$. \end{IEEEproof}

\vspace{3mm}
These latter results allow us to extract the distribution of the respective minimum horizontal distance $Z_A\blank$, as shown in what follows.

\begin{theorem} \label{thm:F_Z_A}
Let $N_A$ denote the number of deployed UAV-mounted BSs, then the CDF of the closest horizontal distance to a UAV is \cite{afshang17bpp}
\begin{align}
F_{Z_A}(z)= 1-\Bar{F}_{\Omega_A}^{N_A}(z)\blank.
\end{align}
\end{theorem}
\begin{IEEEproof}
  Since $Z_A=\min\limits_i\{\Omega_{A,i}\}$, then we can derive its CDF as
  \begin{align}F_{Z_A}(z)=&\;\mathbb{P}(Z\leq z)=1-\mathbb{P}\big(\min\limits_i \{\Omega_{A,i}\}>z\big) 
  =1-\Bar{F}_{\Omega_A}^{N_A}(z).\nonumber \end{align}
\end{IEEEproof}

\begin{corollary} \label{cor:f_Z_A}
  The PDF of the closest ground distance to a UAV is 
\begin{align} 
    f_{Z_A}(z) = N_A\,\Bar{F}_{\Omega_A}^{N_A-1} (z)
    \,f_{\Omega_A}(z)\blank.
\label{eq:f_Z_A}
\end{align}
\end{corollary}
\begin{IEEEproof}
The result trivially follows from taking the derivative of $F_{Z_A}(z)$ with respect to $z$.
\end{IEEEproof}

\subsection{Association Probabilities}
The $B$-association probability quantifies the likelihood that the UE associates to an $B$-BS. 
Based on our assumptions, this can be computed as the probability that the maximum average received power comes from the closest $B$-BS, as conveyed in the following theorem. 

   \begin{table*}[t]
  \begin{center}
    \caption{Minimum Interferer Distances $\mathcal{D}_{BC}(z)$}
    \vspace{-3mm}
    \label{tab:Distances}
    \setlength\tabcolsep{6pt}
     \begin{tabular}{|c|c|c|c|c|} \hline
      \diagbox{\boldmath{$B$}}{\boldmath{$C$}} & \boldmath{$T$} & \boldmath{$A$} \\
      \hline
      \boldmath{$T$} & $z$ 
      & $\begin{cases}
      \frac{\xi_A}{\xi_T}^\frac{1}{\alpha_A}\,z^\frac{\alpha_T}{\alpha_A}, \, \normalfont\text{ if } z>\mathcal{D}_{AT}(0) \nonumber\\
      h\blank, \, \normalfont\text{ otherwise}
      \end{cases}$ \\ \hline
      \boldmath{$A$} & $\frac{\xi_T}{\xi_A}^\frac{1}{\alpha_T}\,(\sqrt{z^2+h^2})^\frac{\alpha_A}{\alpha_T}$
      & $\sqrt{z^2+h^2}$ \\ \hline
    \end{tabular}  
  \end{center} 
\end{table*}

\begin{figure}
\centering
\includegraphics[width=0.7\columnwidth, trim={10.4cm 0.5cm 9.5cm 12cm},clip]{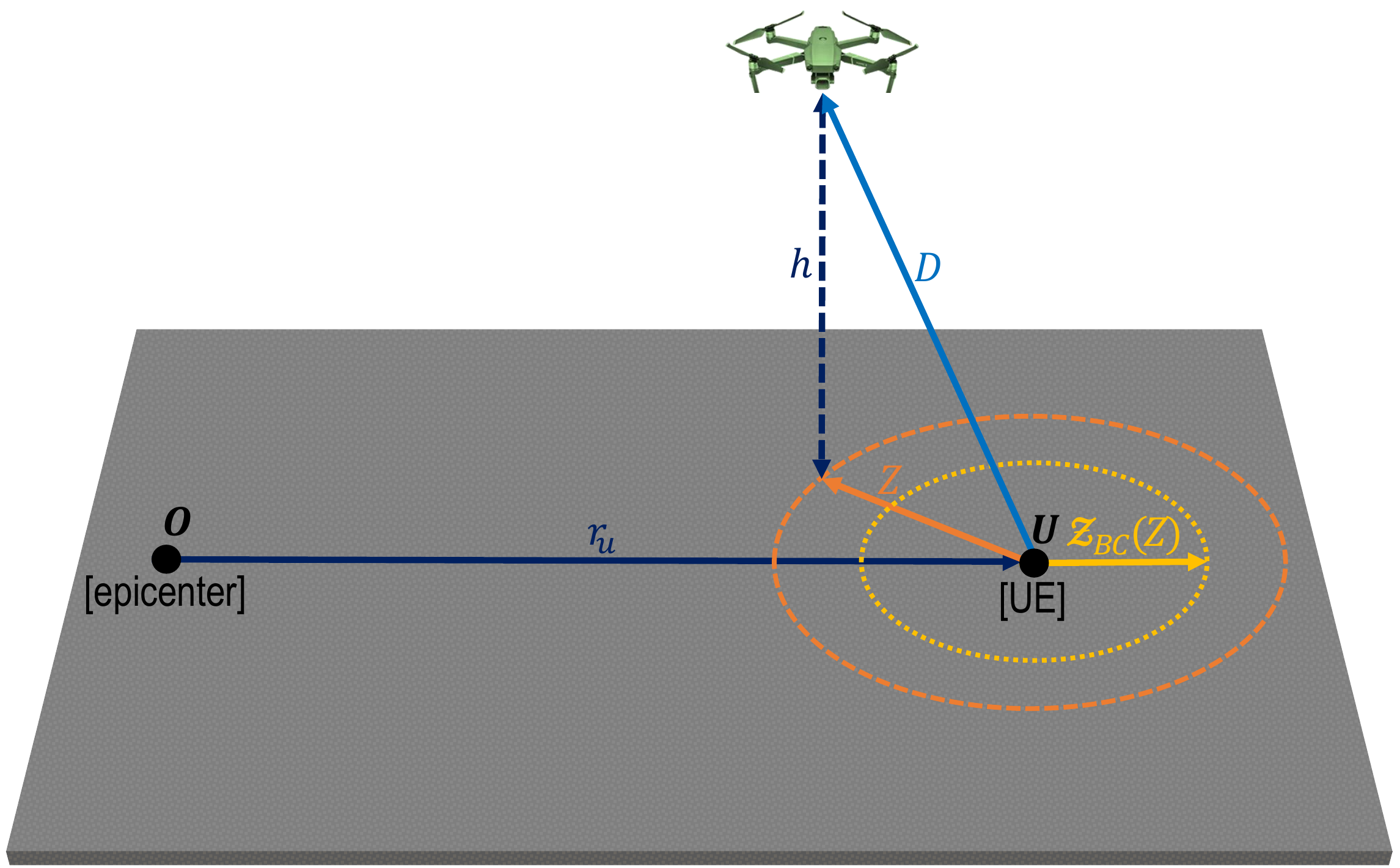}
\vspace{-5mm}
\caption{Generic representation of the minimum interferer horizontal distance $\mathcal{Z}_{BC}(Z)$ in relation to the Euclidean and horizontal distances to the closest $B$-BS.
Note also that, for the sake of an easier representation, the minimum interferer distance $\mathcal{D}_{BC}(Z)$ has been omitted.}
\label{fig:distances}
\end{figure}

\begin{theorem} \label{thm:assoc}
Recalling the subscripts defined in Table~\ref{tab:subscripts}, we denote as $\mathcal{D}_{BC}(z)$ the minimum Euclidean distance of any $C$-interferer if the user associates to a $B$-BS situated at ground distance $z$.
Consequently, $\mathcal{Z}_{BC}(z)=
\begin{cases}
\sqrt{\mathcal{D}_{BC}^2(z)-h^2},\,&\normalfont\text{if }C=A\\
\mathcal{D}_{BC}(z),\,&\normalfont\text{if }C=T\end{cases}$ expresses the horizontal projection of $\mathcal{D}_{BC}(z)$ (see Fig.~\ref{fig:distances}).
Let $r^{\bm\pm} = r_d\pm r_u\blank$, $\mathscr{R}_T=\big[0,\infty\big[\;$, and  $\mathscr{R}_A=\big[\max(0,-r^{\bm-}),r^{\bm+}\big]\blank$,\footnote{$\,$It is evident that the first argument of $\max(x,y)$ is chosen if and only if the user is located inside $\mathscr{A}_d\blank$.} then each $B$-association probability can be expressed as
\begin{align}
\mathcal{A}_B=& \integral_{\mathscr{R}_B} f_{Z_B}(z)\,a_B(z)\,{\rm d}{z}\blank,
\end{align}
where $a_B(z)=\prod\limits_{C\neq B}\Bar{F}_{Z_C}\big(\mathcal{Z}_{BC}(z)\big)$ represents the association probability conditioned on the association to a $B$-BS, which we refer to as the conditional $B$-association probability. 
   \end{theorem}
 \begin{IEEEproof}See Appendix~\ref{appx:assoc}. \end{IEEEproof}
 
\subsection{Conditional Laplace Transforms of the Interference}
Assuming that all BSs operate within the same frequency band, it follows that co-channel interference is generated by each BS except the tagged one. 
Therefore, it is possible to characterize the interference statistics by computing the Laplace transform of the RV $I$, which denotes the aggregate interference.
To this extent, the theorems in this subsection preliminary provide the expressions of the conditional Laplace transforms of the interference generated by each tier of base stations (BSs). \par
Again, the first result we propose refers to the working TBSs, now considered as interferers in the following theorem.

\begin{theorem}\label{thm:Lap_BT}
By recalling the expression of $r_\Omega(\omega,\beta)$ from Theorem~\ref{thm:CDF_T}, the conditional Laplace transform of the interference due to TBSs can be expressed as
\begin{align} 
\mathcal{L}_{I_{BT}}(s|z) =&\; \exp\bigg(-\integral_0^{2\pi}\! \integral_{\mathcal{Z}_{BT}(z)}^\infty\!\! \lambda_T\big(r_\Omega(\Check{\omega},\beta)\big)\,
\mathcal{I}_T(s|\Check{\omega})\,\Check{\omega}\,{\rm d}\Check{\omega}\,{\rm d}\beta\bigg), 
\end{align}
where $\mathcal{I}_T(s|\omega)=1-\Big(\frac{m_T}{m_T\,+\,\xi_T\,s\,\omega^{-\alpha_T}}\Big)^{m_T}$.
\end{theorem}
\begin{IEEEproof}See Appendix~\ref{appx:Lap_BT}. \end{IEEEproof}

\vspace{4mm}
Again, due to the greater complexity of the BPP compared to the PPP, an intermediate step is required to obtain the conditional Laplace transform of the interference coming from the aerial nodes.
To this extent, the following lemma defines the expression of the PDF of the horizontal distance $\Check{\Omega}_A$ between the UE and any interfering ABS.

\begin{lemma}
Let $r^{\bm+} = r_d+r_u\blank$, then the aerial interferers' horizontal distances $\Check{\Omega}_{A,i}$'s constitute an unordered set of iid RVs with PDF expressed as
\begin{align} 
    \;\;\;f_{\Check{\Omega}_{A}}(\Check{\omega} | z) = \frac{f_{\Omega_A}(\Check{\omega})}{\Bar{F}_{\Omega_A}(z)}\blank, \, z\leq \Check{\omega}\leq r^{\bm+}. 
\label{eq:f_checkOmega_A}
\end{align}
\end{lemma}
\begin{IEEEproof}
Let us preliminarily define $n_1=1+\ind(B=A)$ and $\Check{N}_A=N_A-\ind(B\!=\!A)$.
Then, the conditional joint PDF of the aerial interferers' horizontal distances is
  \begin{align}
      f_{\Check{\Omega}_{BA,i}}(\Check{\omega}_{n_1},...,\Check{\omega}_{N_A} | z) 
      \overset{(a)}{=}&
      \frac{N_A!\,f_{\Omega_A}(z)\!
      \prod\limits_{i=n_1}^{N_A} \! f_{\Omega_A}(\Check{\omega}_i)}{f_{Z_A}(z)\,\Bar{F}_{\Omega_A}^{\ind(B\neq A)}(z)} 
      \overset{(b)}{=} \Check{N}_A!\!\prod\limits_{i=n_1}^{N_A} \! \frac{f_{\Omega_A}(\Check{\omega}_i)}{\Bar{F}_{\Omega_A}(z)}\,,
  \end{align}
  where $(a)$ follows from the joint PDF for the order statistics of a sample of size $N_A$ drawn from the distribution of $\Omega_A$ \cite[Appendix C]{Chetlur3D_bpp}, and (b) follows by expressing the term $f_{Z_A}(z)$ as in (\ref{eq:f_Z_A}).
  By recalling \cite[Lemma 3]{afshang17bpp}, we notice that the factorial term $(N_A-1)!$ indicates all possible permutations of the elements in the ordered set of the aerial interferers' horizontal distances.
  As a result, by the joint PDF of the ground distances in the ordered set, the corresponding ground distances in the unordered set are iid with PDF given by (\ref{eq:f_checkOmega_A}).
\end{IEEEproof}

\vspace{4mm}
As already anticipated, we can now express the conditional Laplace transform of the aerial interference by means of the following theorem.

\begin{theorem}\label{thm:Lap_BA}
The conditional Laplace transform of the interference due to the ABSs in the case of $B$-association can be expressed as 
\begin{align} 
\mathcal{L}_{I_{BA}}(s|z) = \Check{\Upsilon}_{BA}^{\Check{N}_A}(s,z) , 
\label{eq:LapBA} \end{align}
where $\Check{\Upsilon}_{BA}(s,z)=\!\integral_{\mathcal{Z}_{BA}(z)}^{r^{\bm+}} \mathcal I_A(s|\Check{\omega})\,f_{\Check{\Omega}_A}\big (\Check{\omega}|\mathcal{Z}_{BA}(z)\big)\,{\rm d}\Check{\omega}$
with
$\mathcal I_A(s|\Check{\omega})=\Big(\frac{m_A}{m_A\,+\,\xi_A\,s\,\mathcal{D}_{AA}^{-\alpha_A}(\Check{\omega})}\Big)^{m_A}$.
\end{theorem}
\begin{IEEEproof}
See Appendix~\ref{appx:Lap_BA}.
\end{IEEEproof}

\vspace{4mm}
To conclude, the following corollary defines the conditional Laplace transform of the aggregate interference.

\begin{corollary}
The conditional Laplace transform of the aggregate interference can be expressed as 
\begin{align}
\mathcal{L}_{I,B}(s|z)=&\prod\limits_{C=\{A,T\}} \mathcal{L}_{I_{BC}}(s|z)\blank.
\end{align}
\end{corollary}
\begin{IEEEproof}
The proof trivially follows by recalling that the aggregate interference is the sum of the interferences coming from each layer. 
\end{IEEEproof}

\subsection{Coverage Probability}
Based on the expressions derived for the PDFs of the distance to the closest BS, the conditional association probabilities, the $\rm SINR$, and the Laplace transform of the interference, we hereby provide the exact and approximate expressions of the coverage probability under Nakagami-\it{m} fading conditions.

\begin{theorem}\label{thm:Pc}
Let $p_{c,B}(z)$ denote the exact coverage probability conditioned on the association to a $B$-BS located at horizontal distance $z$ within its own planar domain $\mathscr{R}_B$ defined as in Theorem~\ref{thm:assoc}.
Then, the exact coverage probability for a typical user in the wireless system described in Section~\ref{sec:SystemModel} is given by
\begin{align}
     {P}_c=&\!\sum\limits_{B=\{A,T\}}\, \integral_{\mathscr{R}_B} a_B(z) \, p_{c,B}(z) \, f_{Z_B}(z)\,{\rm d}{z}\blank,
     \end{align}
where 
\begin{align}p_{c,B}(z)=\sum\limits_{k=0}^{m_B-1} \frac{\big(-\mu_B(z)\big)^k}{k!}\,\frac{\partial^k}{\partial s^k} \mathcal{L}_{J,B}(s|z)\Big|_{s=\mu_B(z)}
\end{align}
with $\mathcal{L}_{J,B}(s|z)=e^{-s\,\sigma_n^2}\,\mathcal{L}_{I,B}(s|z)$ and $\mu_B(z)=m_B\,\frac{\tau}{\xi_B}\,\mathcal{D}_{BB}^{\alpha_B}(z)$.
The expressions of the functions $f_{Z_B}(z)$'s are provided in Corollaries~\ref{cor:PDF_T} and~\ref{cor:f_Z_A}; the general expression of the association probabilities $a_B(z)$'s is given by Theorem~\ref{thm:assoc}; the functions $\mathcal{L}_{I,B}(s|z)$'s respectively refer to Theorems~\ref{thm:Lap_BT} and~\ref{thm:Lap_BA}.
\end{theorem}
\begin{IEEEproof}
  See Appendix~\ref{appx:Pc}.
\end{IEEEproof}

\vspace{4mm}
Since computing the exact expression of the conditional coverage probability may require computing high-order derivatives of the conditional Laplace transform of the interference, it is usually preferable to approximate it, as suggested by the following theorem.

\begin{theorem}\label{thm:Pc_tilde}
To ease the evaluation of the coverage probability, the conditional coverage probability can be approximated as~\cite[Sec. III-D]{Alzenad19}
\begin{align} \label{eq:approx_PcB}
\bm\tilde{p}_{c,B}(z)=\sum\limits_{k=1}^{m_B}
\binom{m_B}{k}\,(-1)^{k+1}\,\mathcal{L}_{J,B}\big(k\,\varepsilon_{2,B}\,\mu_B(z),z\big)\blank,
\end{align}
where $\mathcal{L}_{J,B}(s|z)$ and $\mu_B(z)$ are given by Theorem~\ref{thm:Pc}, and $\varepsilon_{2,B}=(m_B!)^{-\frac{1}{m_B}}$.
\end{theorem}
\begin{IEEEproof}See Appendix~\ref{appx:Pc_tilde}. \end{IEEEproof}
  
\section{Results and Discussion}
In this section, the analytical results based on the expressions derived in Sec.~\ref{sec:PerformanceAnalysis} are verified by means of Monte Carlo simulations.
By inspecting these results, we will try to understand how each system parameter affects the network's performance { as defined in Theorem~\ref{thm:Pc_tilde} (or Theorem~\ref{thm:Pc} when no ABSs are deployed)}. 
However, let us recall that we are: \textit{(i)} assuming ideal backhaul links, \textit{(ii)} evaluating the QoS only in terms of coverage probability, and \textit{(iii)} for the sake of conciseness and mathematical tractability, not specifying the difference between LoS and NLoS transmissions for both the terrestrial and aerial tiers (this, however, is a precautionary assumption since the presence of NLoS nodes would strongly reduce the average aerial interference power without significantly increase the average desired signal power, leading to more optimistic behaviors of the coverage probability as { the number of UAVs} $N_A$ is increased).
Note also that aerial BSs present limitations in terms of capacity (\it{e.g.}, due to the small number of antennas supported) and autonomy, which considerations are beyond the scope of this study.
Nonetheless, the observed trends can effectively support cellular operators in network planning. 
For example, they would be able to quantify the advantage of strengthening the terrestrial infrastructure, as well as predicting the number of ad hoc ABSs needed in the occurrence of a specific calamity.\par
For this study, unless stated otherwise, we have assumed the values of the parameters according to Table~\ref{tab:Parameters} and used markers and lines to represent analysis and simulation results, respectively.
Note also that, although the standard values of { the disaster radius} $r_u$ and { the QoR} $\chi$ do not reflect typical disaster scenarios (since the majority of the users should be close to the edge of { the disaster-struck area} $\mathscr{A}_d$ and the network should not be fully destroyed), they may correspond to the most critical situation (since users at the disaster epicenter have more chances to be trapped and/or seriously injured, { and a less resilient network has more chances of becoming overloaded}).

 \begin{table*}[t] 
  \begin{center}
    \caption{Main system parameters' standard values}
    \label{tab:Parameters}
        \vspace{-3mm}
    \begin{tabular}{|c|c|} 
    \hline
      \textbf{Parameters} &
      \textbf{Values} \\\hline
      Original TBSs' density & $\lambda_0=3\,$TBSs/km$^2=3\!\times\!10^{-6}\,$TBSs/m$^2$ \\\hline
       ABSs' altitudes &
        $h=[0.2,\,20]\,\text{km}=[200,\,2\!\times\! 10^4]\,\text{m, respectively for LAPs and HAPs}$ \\\hline
       QoR & $\chi=0$ \\\hline
       UEs' distance from the origin & $r_u=0$ \\\hline
       Disaster radius & $r_d=[1,\,10]\,\text{km, respectively for LAPs- and HAPs-based solution}$ \\\hline
       Path loss exponents &
       $\begin{cases}\alpha_A=[3,\,2.5]\text{, respectively for LAPs and HAPs}\\
       \alpha_T=3.5 \end{cases}$ \\\hline
       Transmit powers &
       $\begin{cases}\rho_A=[5,\, 50]\text{ W, respectively for LAPs and HAPs}\\
       \rho_T=10\text{ W} \end{cases}$ \\\hline
       Nakagami-\it{m} shape parameters &
       $m_A=2 ,\; m_T=1$ \\\hline
       SINR threshold &
       $\tau=-5\,\text{dB}=0.3162$\\ \hline
       Noise power &
       $\sigma_n^2=10^{-12}\,\text{W}$ \\\hline
       Mean additional transmit losses &
       $\eta_A=-1.6\,\text{dB}=0.6918 ,\; \eta_T=-2\,\text{dB}=0.631$  \\\hline
     \end{tabular}  
  \end{center}
\end{table*} 

\subsection{Influence of the Disaster Radius}

\begin{figure}
\centering
{\includegraphics[width=0.6\columnwidth, trim={0.7cm 0cm 1.5cm 0cm},clip]{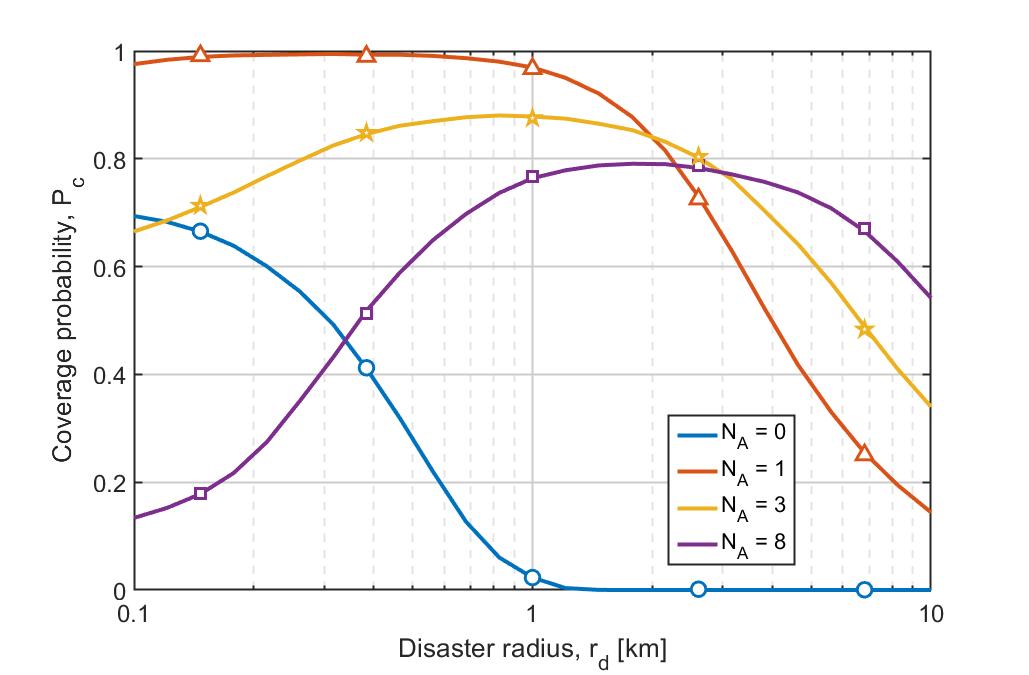}} \\
\vspace{-1mm}
(a) \\
{\includegraphics[width=0.6\columnwidth, trim={0.7cm 0cm 1.5cm 0cm},clip]{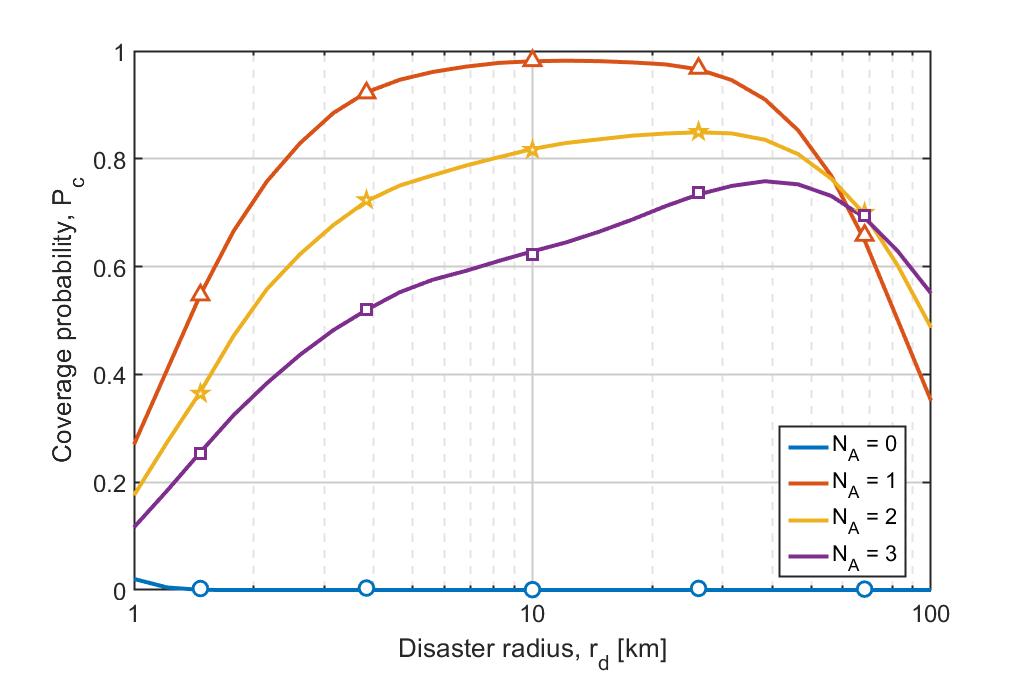}} \\
\vspace{-1mm}
(b)\\
\caption{Coverage probability as function of the disaster radius when the ABSs deployed are: (a) LAPs and (b) HAPs.}
\label{fig:Pc_VS_Rd}
\end{figure}

In Fig.~\ref{fig:Pc_VS_Rd}, the impact of { the disaster radius} $r_d$ is investigated under the assumption that either LAPs, HAPs, or none of them are deployed.
In particular, the plots describe different coverage probability behaviors depending on the cardinality $N_A$ of each type of fleet of ABSs.
Different positive values of $N_A$ have been selected for LAPs and HAPs because of their difference in terms of coverage area.
Finally note that, to better highlight the influence of $r_d\blank$, here we assume $\chi=0$ and $r_u=0\blank$. 
From Fig.~\ref{fig:Pc_VS_Rd} we can extract some precious insights:

\subsubsection{Outer TBSs} Based on the considered system parameters, outer TBSs can support the UE (assuming $r_u=0$) only for very small values of $r_d\,$;
in other words, $P_c$ rapidly approaches zero as $r_d$ exceeds a couple of hundred meters.
{ This occurs because a larger disaster radius implies a longer average distance to the closest functioning TBS (which in this case is lower-bounded by $r_u$):
in other words, the higher path loss overcompensates the weaker interference. }
Therefore, unless $\mathscr{A}_d$ is very small, non-resilient networks { ($\chi=0$)} should not be considered self-sufficient. \par

\subsubsection{LAPs} If $r_d$ is less than two kilometers, deploying one single LAP is generally the optimal choice, as the red curve in Fig.~\ref{fig:Pc_VS_Rd}-a confirms.
Our explanation to this fact relies in the well-known trade-off for VHetNets' densification: while increasing the number of nodes statistically reduces the distance to the tagged BS, it increases the power of the aggregate interference. 
However, for $r_d\geq2\,$km the optimal $N_A$ rapidly increases: for $r_d=10\,$km even eight LAPs are not enough to ensure sufficient reliability (for which we may expect $P_c\gtrapprox0.6$) at the epicenter of the calamity. 

\subsubsection{HAPs} In case of relatively small disasters, deploying HAPs is highly discouraged, as Fig.~\ref{fig:Pc_VS_Rd}-b confirms.
On the other side, as $r_d$ exceeds a few kilometers, HAPs can be successfully deployed by leveraging their strong transmit power and favorable channel conditions. 
The optimal cardinality of the fleet is usually $N_A=1$, but it rapidly increases as $r_d\to100\,$km: in fact, here the aerial interference experienced at the disaster epicenter becomes much less detrimental while a higher value of $N_A$ generally implies a shorter distance between the UE and the closest HAP.

\subsection{Influence of the UE's Location}

\begin{figure}
\centering
{\includegraphics[width=0.6\columnwidth, trim={0.7cm 0cm 1.5cm 0cm},clip]{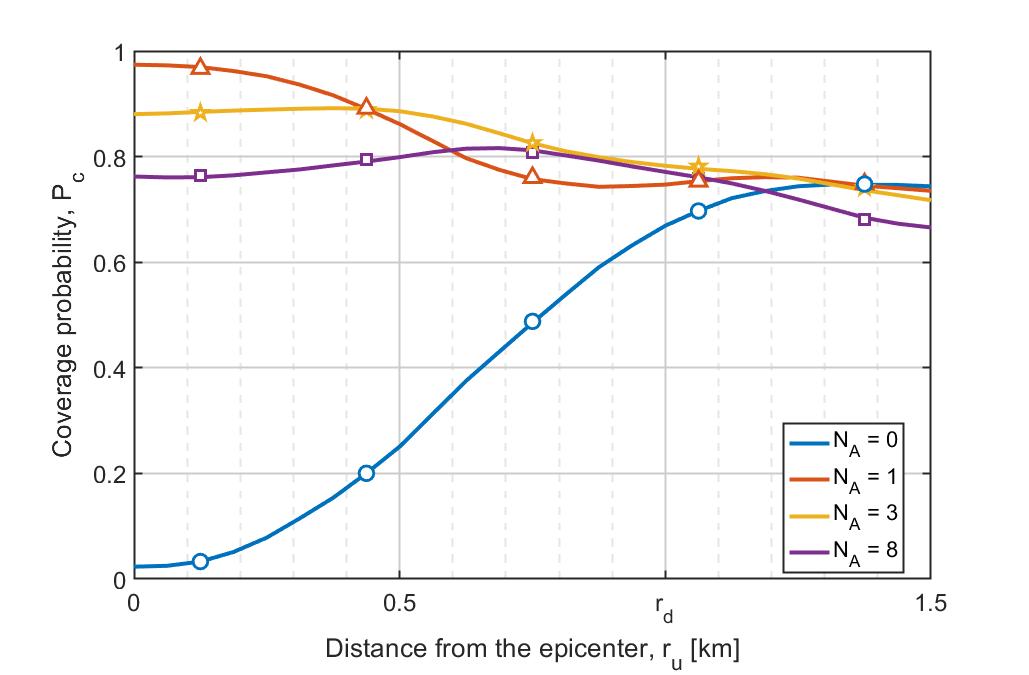}} \\
\vspace{-1mm}
(a) \\
{\includegraphics[width=0.6\columnwidth, trim={0.7cm 0cm 1.5cm 0cm},clip]{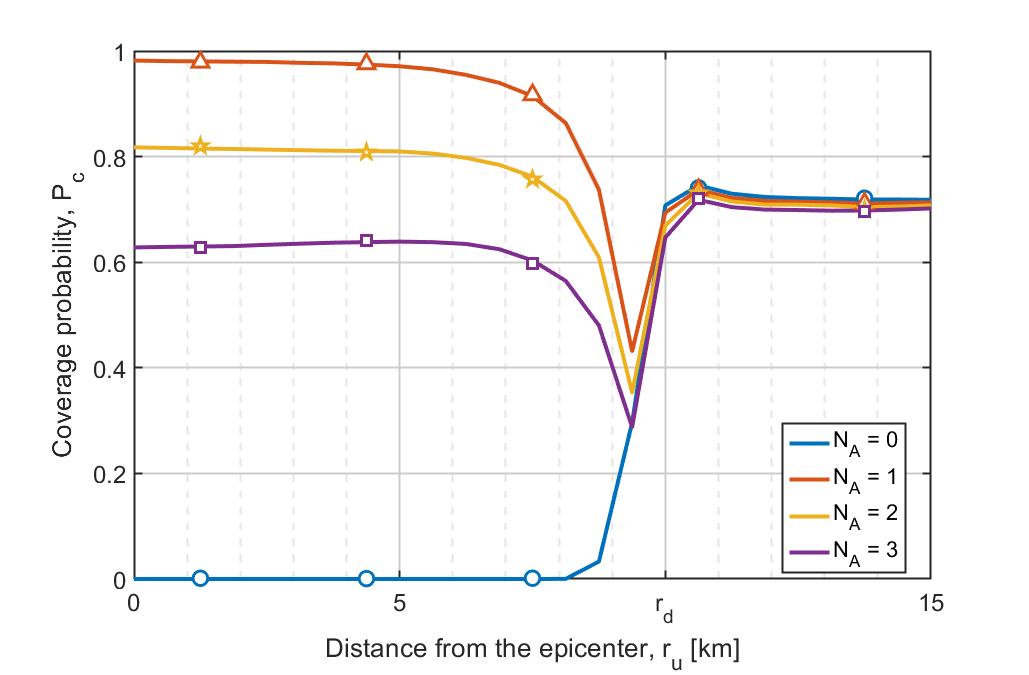}} \\
\vspace{-1mm}
(b) \\
\caption{Coverage probability as function of the UE's location when the ABSs deployed are: (a) LAPs and (b) HAPs.
The disaster radius is assumed equal to one and ten kilometers, respectively.}
\label{fig:Pc_VS_Ru}
\end{figure}

In this subsection we investigate the influence, in terms of coverage probability, of the distance of the user with respect to the epicenter.
This time, $r_d$ is specifically fixed in order to simulate a typical disaster scenario for various LAPs- or HAPs-aided networks, which based on the results obtained in Fig.~\ref{fig:Pc_VS_Rd} (and also in our previous paper~\cite{Matracia21disaster}) are assumed to be conveniently deployed in case of small or large disasters, respectively. 
Once again, we assume there are no surviving TBSs inside the disaster-struck zone. 

\subsubsection{Outer TBSs} As expected, the blue curves in Fig.~\ref{fig:Pc_VS_Ru} convey that the TBSs surrounding $\mathscr{A}_d$ can be quite effective in serving UEs located relatively close to the edge of the suffered region (roughly within a hundred meters). 
We can also see that the QoS experienced by the typical user slightly depends on $r_d\blank$, and is mostly affected by the distance to the closest working TBS.  

\subsubsection{LAPs} Fig.~\ref{fig:Pc_VS_Ru}-a illustrates an overall improvement when deploying low-altitude aerial nodes above a relatively small disaster region of radius $1\,$km.
We can notice that all the considered LAP fleets are able to cover more than twice the area of $\mathscr{A}_d\blank$.
In addition, for a typical user located at the origin the highest QoS is achieved for $N_A=1\blank$, as anticipated in Fig.~\ref{fig:Pc_VS_Rd}-a.\\
For the considered setup, a high number of aerial nodes (see the violet curve) would not maximize the network performance for any distance from the epicenter, and therefore is not recommended.
Finally, for outer UEs the aerial interference is quite negligible as long as $N_A\leq3\blank$, which would promote the deployment of multiple LAPs in case of a considerable traffic demand. 

\subsubsection{HAPs} For this scenario, we considered a disaster radius of $10\,$km.
From Fig.~\ref{fig:Pc_VS_Ru}-b, it is evident that deploying multiple HAPs is not convenient since it implies a strong aerial interference, although it might be needed in case of a considerably larger size of the disaster.\\
Furthermore, we can state that the users located around the epicenter are the ones which benefit the most from the deployed HAPs, up to the point that for $N_A\leq2$ their experienced post-disaster QoS surpasses its pre-disaster counterpart.
As the typical user moves away from the epicenter, the $A$-association generally decreases and the aerial interference becomes more and more problematic; then, a minimum coverage is experienced at around $600\,$m before the edge of $\mathscr{A}_d\blank$, where the outer TBSs become close enough to frequently serve the user and the terrestrial interference dominates over its aerial counterpart.

\subsection{Influence of the QoR}
Let us now focus on the concept of resiliency by evaluating the coverage probability for various expressions of $\chi(r)$.
We propose two studies to better understand the importance of having resilient TBSs: one assumes a uniform QoR and includes ABSs in the network architecture, whereas the other investigates various QoR's planar distributions while omitting ABSs. 

\subsubsection{Uniform QoR} 

\begin{figure}
\centering
{\includegraphics[width=0.6\columnwidth, trim={0.75cm 0cm 1.55cm 0cm},clip]{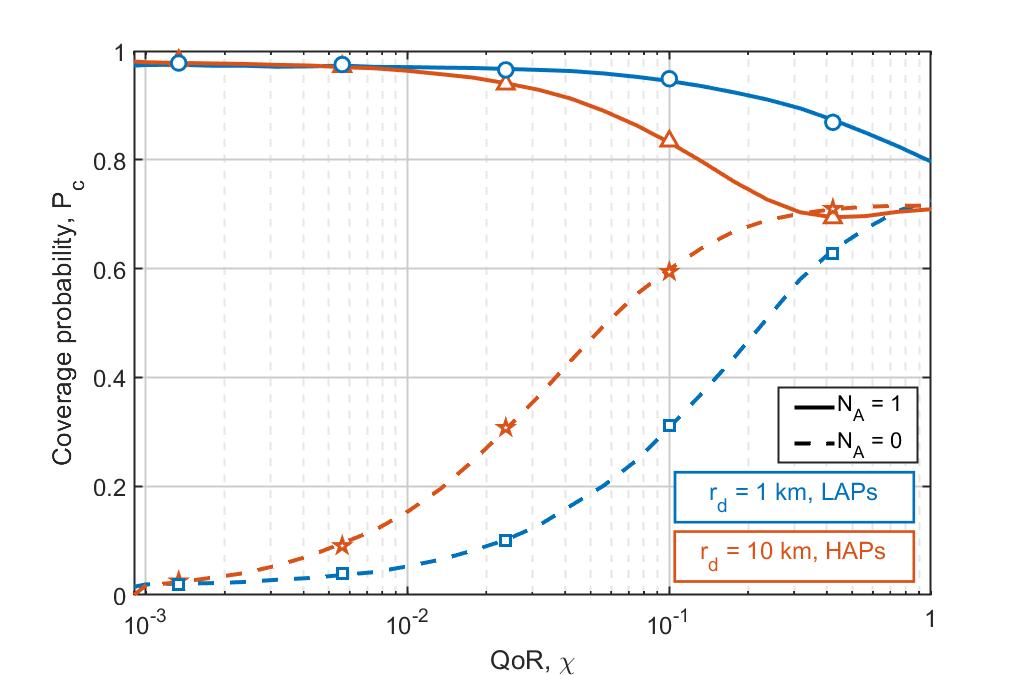}} 
\vspace{-1mm}
\caption{Coverage probability as a function of the QoR (assumed uniform over the disaster-struck area) and the user is located at the origin.}
\label{fig:Pc_VS_chi}
\end{figure}

The dashed curves of Fig.~\ref{fig:Pc_VS_chi} tell us that, from a pure coverage perspective, if $N_A=0$ then a small value of $\mathscr{A}_d$ can be much more problematic than a hundred times larger one.
This can be explained by taking into account that a larger disaster area paradoxically benefits a typical user located at its epicenter because it increases the distance to the interfering outer TBSs.
Instead, the solid lines show that even one single aerial node can lead to a high $A$-association probability (because of the advantaged channel conditions) and consequently boost the coverage probability, especially as $\chi\to0\blank$, which confirms the results previously shown in Figs.~\ref{fig:Pc_VS_Rd} and~\ref{fig:Pc_VS_Ru}. \\
Generally speaking, $P_c$ decreases or increases depending on whether the ABS (and especially the HAP) is present or not.
As Fig.~\ref{fig:Pc_VS_chi} illustrates, by assuming a fully-resilient terrestrial network (which is equivalent to the scenario without any perturbation) we would always have $0.7<P_c<0.8$, meaning that the aerial node would not remarkably improve the existing infrastructure.
Actually, for $\chi\geq0.4\,$ a small degradation of $P_c$ due to the presence of a HAP can be observed by comparing the red curves. 
\subsubsection{QoR Distributions} 

\begin{figure}
\centering
{\includegraphics[width=0.6\columnwidth, trim={0.7cm 0cm 1.45cm 0cm},clip]{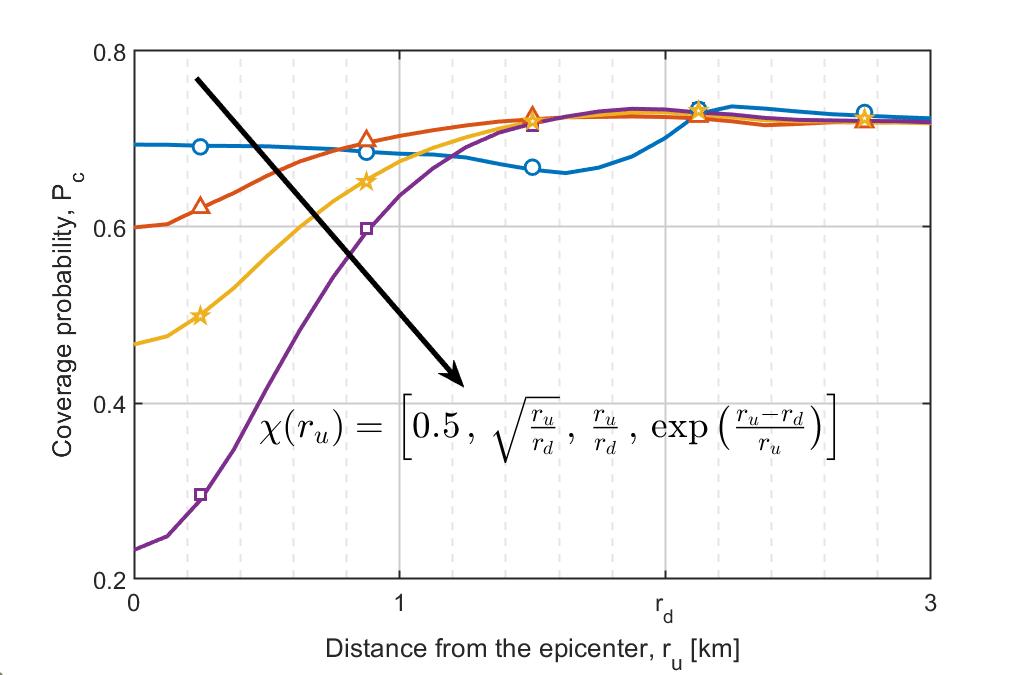}}
\vspace{-1mm}
\caption{Coverage probability as function of the UE's location for various QoR behaviors, considering $N_A=0$ and $r_d=2\,$km.}
\label{fig:Pc_VS_chi(r)}
\end{figure}

We hereby consider a medium-size disaster with $r_d=2\,$km and investigate the behavior of $P_c$ as a function of $r_u\blank$.
As a parameter, we consider various planar distributions (namely, constant, \it{square-root-like}, linear, and exponential with respect to $r_u$) of the QoR and compare them in the presence of only terrestrial nodes.
The expression of $\chi(r_u)$ might strongly depend on the entity of the disaster: for example, we may expect an explosion leading to a sharp variation of the density of surviving TBSs as we move away from the origin, while a flood should have a more uniform influence on the surrounding environment. \\
All the curves displayed in Fig.~\ref{fig:Pc_VS_chi(r)} convey that, compared to the case with full TBS density (which can be fairly assumed if $r_u\gg r_d$), even halving the original TBS density (that is, reducing { the QoR} to just $0.5$) would nowhere compromise the QoS, because $r_d$ is relatively small.
Consistently with Fig.~\ref{fig:Pc_VS_chi}, we can also notice that for $r_u\to0$ the QoS directly depends on the local density $\lambda_T(r_u)$;
in this scenario, the interference coming from the outer TBSs is usually negligible and the inner interferers are much farther than the tagged TBS. 
Instead, if the user is close to the edge of $\mathscr{A}_d\blank$, the interference becomes relevant and hence a higher density of the surrounding TBSs leads to a slightly lower value of $P_c\blank$. 

\section{Conclusion and Future Works} \label{sec:Conclusion}
In this paper, we proposed a concise and tractable mathematical framework which borrows tools from SG and makes use of indicator functions in order to enable the estimation of the QoS in UAV-assisted post-disaster wireless networks.
In particular, given a typical VHetNet consisting of partially-resilient TBSs and ad-hoc ABS, we provided novel analytical expressions for the minimum distance distributions, association probabilities, and Laplace transforms of the interference in order to obtain the exact and approximate expressions of the coverage probability experienced by a typical UE, which can be arbitrarily located anywhere on the ground plane. 
Furthermore, by verifying the obtained numerical results, we proved that a properly chosen fleet of ad-hoc ABSs can strongly support the terrestrial infrastructure in various scenarios, and highlighted the trade-off between wider coverage and stronger interference due to heterogeneous network's densification. \par 
This study could be extended in various research directions.
For example, a more general setup where the ABSs operate in either LoS or NLoS condition with respect to the user should be considered in the future.
Furthermore, it would be interesting to evaluate novel solutions for interference mitigation in post-disaster scenarios, perhaps by switching off some specific extra-region TBSs that are unlikely to serve any high-priority users involved in the disaster; this strategy would also help to reduce the overall power consumption, which is a critical issue since power systems are also susceptible to calamities.
Finally, other important aspects such as network overload and backhaul issues could be taken into account in future extensions of this work.

\appendices
\section{Proof of Theorem~\ref{thm:CDF_T}}\label{appx:CDF_T}
Recalling that the altitude of terrestrial antennas is assumed to be negligible compared to that of their aerial counterparts, the ground (and Euclidean) distance between the UE and the tagged TBS is identified by the RV $Z_T$.
Thus, the expression of the respective CDF can be derived from the null probability of the PPP \cite{Andrews11}. 
Let $N_T(z)$ be the number of TBSs residing within a distance $z$ from the UE, then:
\begin{align}    
F_{Z_T}(z)=&\;\P(Z_T\leq z)=1-\P(Z_T>z)
=1-\P(N_T(z)=0) \nonumber\\
=&\;1 - \exp\bigg( -\integral_0^{2\pi} \integral_0^z \lambda_T\big(r_\Omega(\omega,\beta)\big)\,\omega\,{\rm d}\omega\,{\rm d}\beta\bigg), 
\end{align}
where $r_\Omega(\omega,\beta)=\sqrt{r_u^2+\omega^2-2\,r_u\,\omega\,\cos\beta}$ and $\lambda_T(\omega,\beta)$  describe the horizontal distance from the origin and the behavior of the post-disaster TBSs' density, respectively. 

\section{Proof of Corollary~\ref{cor:PDF_T}}\label{appx:PDF_T}
\it{The derivative of $F_{Z_T}(z)$ is 
\begin{align}
    f_{Z_T}(z) =& -\exp\!\Bigg( -\integral_0^{2\pi} \integral_0^z \lambda_T\big(r_\Omega(\omega,\beta)\big)\,\omega\,{\rm d}\omega\,{\rm d}\beta\Bigg)\,\Bigg(-\frac{{\rm d}}{{\rm d}z} \integral_0^{2\pi} \integral_0^z \lambda_T\big(r_\Omega(\omega,\beta)\big)\,\omega\,{\rm d}\omega\,{\rm d}\beta\Bigg),
\end{align}
where, introducing $g_z(z,\beta) = \integral_0^z \lambda_T\big(r_\Omega(\omega,\beta)\big)\,\omega\,{\rm d}\omega$ and applying the Leibniz integral rule, we have
\begin{align} 
    \frac{{\rm d}}{{\rm d}z} \integral_0^{2\pi} g_z(z,\beta)\,{\rm d}\beta =&\;
    g_z(z,2\pi)\,\frac{{\rm d}}{{\rm d}z}(2\pi)-g_z(z,2\pi)\,\frac{{\rm d}}{{\rm d}z}(0)+ \integral_0^{2\pi} \frac{\partial}{\partial z}\,g_z(z,\beta)\,{\rm d}\beta 
    =\integral_0^{2\pi} \frac{\partial}{\partial z}g_z(z,\beta)\,{\rm d}\beta \nonumber\\
    =&\; z \integral_0^{2\pi} \lambda_T\big(r_\Omega(z,\beta)\big)\,{\rm d}\beta\blank,
\label{eq:dz_02pi} \end{align}
which completes the proof.}

\section{Proof of Theorem~\ref{thm:assoc}} \label{appx:assoc}
The $B$-association probability represents the probability that the UE associates to a BS of type $B$ (\it{i.e.}, the average power received from the closest BS of type $B$ exceeds the average powers received from the closest BSs of the other types). \par
Now, let $\mathcal{D}_{BC}(z)$ express the minimum Euclidean distance of any interfering BS of type $C$ when the UE associates to a BS of type $B$ located at ground distance $z$.
Noting that the Euclidean distance to the tagged BS equals $\sqrt{z^2+h^2}$ if $B\!=\!A$ and $z$ otherwise, we define the projection on the ground of $\mathcal{D}_{BC}(z)$ as: \begin{align}
\mathcal{Z}_{BC}(z)=\begin{cases}
\sqrt{\mathcal{D}_{BC}^2(z)-h^2}\blank,&\!\!\normalfont\text{if }C\!=\!A \\
\mathcal{D}_{BC}(z)\blank,&\!\!\normalfont\text{if }C\!=\!T\end{cases}\,\nonumber,
\end{align}
which is conceptually represented in Fig.~\ref{fig:distances}.
\par
By recalling that $\xi_O=\rho_O\,\eta_O$ with $O\in\{A,T\}$, and introducing the random Euclidean distance $D_O=\mathcal{D}_{OO}(Z_O)$ as function of its own horizontal component $Z_O$, we can finally derive the $B$-association probabilities, as follows.

\subsection{$T$-Association Probability}
Recalling that $Q_O^*$ denotes the average power received from the closest $A$- or $T$-BS, then the probability $\P(Q_T^*>Q_A^*)$ depends on $Z_T\blank$, and hence
\begin{align}
\mathcal{A}_T=&\;\P\big(Z_A>\mathcal{Z}_{TA}(Z_T)\big) 
= \integral_0^{\infty} \Bar{F}_{Z_A}\big(\mathcal{Z}_{TA}(z)\big)\,f_{Z_T}(z)\,{\rm d}{z}\blank,
\end{align}
where $a_T(z)=\Bar{F}_{Z_A}\big(\mathcal{Z}_{TA}(z)\big)$ expresses the conditional $T$-association probability. 

\subsection{$A$-Association Probability}
Trivially, the association probabilities are complementary, therefore 
$$\mathcal{A}_A=1-\mathcal{A}_T\blank.$$
 Alternatively, the $A$-association probability can be computed as
 \begin{align}
   \mathcal{A}_A=&\;\P\big(Z_T>\mathcal{Z}_{AT}(Z_A)\big) = \integral_{\max(0,-r^{\bm-})}^{r^{\bm+}} \Bar{F}_{Z_T}\big(\mathcal{Z}_{AT}(z)\big)\, f_{Z_A}(z)\,{\rm d}z\blank.
\end{align}

\section{Proof of Theorem~\ref{thm:Lap_BT}} \label{appx:Lap_BT}
Let us first define the sets of the $C$-BSs as $\Check{\Phi}_C = \Phi_C\backslash\big(\mathscr{B}_z(\mathcal{Z}_{BM}(z))\cup W^* \big)$, where $\mathscr{B}(z)$ denotes the circle of radius $z$ centered around the typical user.
Now, we can denote the set of TBSs' coordinates as {\bf Y} and recall that $\mathcal{I}_T(s|\omega)=1-\Big(\frac{m_T}{m_T\,+\,\xi_T\,s\,\mathcal{D}_{TT}^{-\alpha_T}(\omega)}\Big)^{m_T}$.
In order to obtain the expression of the conditional Laplace transform of the terrestrial interference, we firstly take the expectation over both the point process and the set of fading gains \cite[Sec. III-C]{PRIMER}:
\begin{align}
\mathcal{L}_{I_{BT}}(s|z)=&\;\E\left[e^{-s\,I_{BT}}|z\right]
\overset{(a)}{=}\E_{\Phi_T} \bigg[\prod_{Y_i\in\Check{\Phi}_T} \psi_T(s,Y_i)\bigg]\nonumber\\
\overset{(b)}{=}&\exp\bigg(-\!\!\integral_{\R^2 \backslash \mathscr{B}_z(\mathcal{Z}_{BT}(z))} \hspace{-5mm} \lambda_T(\|{\bf Y}\|)
\,\big(1-\psi_T(s,{\bf Y})\big)\,{\rm d}{\bf Y}\bigg) \nonumber\\
=&\exp\bigg(- \integral_0^{2\pi} \integral_{\mathcal{Z}_{BT}(z)}^\infty \lambda_T\big(r_\Omega(\Check{\omega},\beta)\big)\,\mathcal{I}_T(s|\Check{\omega})\,\Check{\omega}\,{\rm d}\Check{\omega}\,{\rm d}\beta\bigg)\blank.
\end{align}
 Note that $(a)$ follows from the independence of the exponentially distributed gains $G_{T,Y_i}$'s, having introduced the function $\psi_C(s,W_i)=\E_{G_{C,W_i}}\left[\exp\left(-\frac{s\,G_{C,W_i}\,\xi_C}{\|W_i\|^{\alpha_C}}\right)\right]$ for any type of interferers, and $(b)$ derives from the application of the probability generating functional (PGFL) to the latter function. 
 
\section{Proof of Theorem~\ref{thm:Lap_BA}} \label{appx:Lap_BA}
By letting $\Check{\Omega}_A$ denote the set of aerial interferers' horizontal distances and recalling that $\Check{N}_A=N_A-\ind(B\!=\!A)$, the Laplace transform of the aerial interference conditioned on $B$-association can be derived as~\cite[Eq. (4), (16)]{PRIMER}
\begin{align}
\mathcal{L}_{I_{BA}}(s|z) =&\; \mathbb{E}_{I_{BA}}\big[e^{-s\,I_{BA}}|z\big] 
= \mathbb{E}_{I_{BA}} \Big[\exp\Big(-s\,\sum\limits_{i=1}^{\Check{N}_A} G_i\,\mathcal{D}_{AA}^{-\alpha_A}(\Check{\Omega}_{A,i})\Big)\big|z \Big] \nonumber\\
\overset{(a)}{=}&\; \mathbb{E}_{\Check{\Omega}_A} \Big[\mathbb{E}_G \Big[\prod\limits_{i=1}^{\Check{N}_A} \exp\big(-s\,G_i\,\mathcal{D}_{AA}^{-\alpha_A}(\Check{\Omega}_{A,i})\big) \big|z\Big]\Big] \nonumber\\
\overset{(b)}{=}&\; \mathbb{E}_{\Check{\Omega}_A} \Big[ \prod\limits_{i=1}^{\Check{N}_A} \mathbb{E}_{G_i}\big[\exp(-s\,G_i\,\mathcal{D}_{AA}^{-\alpha_A}(\Check{\Omega}_{A,i}))\big]\big|z\Big] \nonumber\\
\overset{(c)}{=}&\; \mathbb{E}_{\Check{\Omega}_A}\Big[\prod\limits_{i=1}^{\Check{N}_A} \mathcal{I}_{A,i}(s|\Check{\Omega}_{A,i}) \big|z\Big] 
\overset{(d)}{=} \bigg( \mathbb{E}_{\Check{\Omega}_{A,i}}\Big[ \mathcal{I}_{A,i}(s|\Check{\Omega}_{A,i})\big|z\Big]\bigg)^{\Check{N}_A},
\end{align}
where $\mathcal I_{A,i}(s|\Check{\Omega}_{A,i})=\Big(\frac{m_A}{m_A\,+\,\xi_A\,s\,\mathcal{D}_{AA}^{-\alpha_A}(\Check{\Omega}_{A,i})}\Big)^{m_A}$. 
Step $(a)$ follows from the independence of the channel gains and the distances of the aerial interferers, whereas $(b)$ follows from rewriting the expectation of a product as a product of the expectations owing to iid channel gains. 
Then, $(c)$ follows from the moment generating function (MGF) of the gamma-distributed fading gains $G_i$'s~\cite[Appendix E]{Chetlur3D_bpp}, and $(d)$ from the conditional iid distances of the aerial interferers.

\section{Proof of Theorem~\ref{thm:Pc}} \label{appx:Pc}
Let us first recall from Table~\ref{tab:Distances} and Theorem~\ref{thm:assoc} the expressions of the Euclidean distances $\mathcal{D}_{BB}(z)$'s and the planar domains $\mathscr{R}_B$'s, respectively.
Now, following the same approach proposed in \cite{Galkin19}, the exact expression of the coverage probability can be obtained as 
\begin{align}
P_c=&\;\E_{Z_B}\big[\P({\rm SINR}>\tau \, | \, Z_B=z)\big] 
=\sum\limits_{B=\{A,T\}} \E_{Z_B}\big[a_B(Z_B)\,p_{c,B}(Z_B)\big]    \nonumber\\ 
=&\sum\limits_{B=\{A,T\}} \, \integral_{\mathscr{R}_B} a_B(z)\,p_{c,B}(z)\,f_{Z_B}(z)\,{\rm d}z\blank,
\end{align}
in which the exact expressions of the conditional coverage probabilities are given by
\footnote{$\,$In the particular case of Rayleigh fading channel ($m_B=1$), we can compute the conditional coverage probability as
$p_{c,B}(z)=\exp\Big({-\frac{\tau\,\mathcal{D}_{BB}^{\alpha_B}(z)\,\sigma_n^2}{\xi_B}}\Big) \,\mathcal{L}_{I,B}\Big(\frac{\tau\,\mathcal{D}_{BB}^{\alpha_B}(z)}{\xi_B},z\Big)\blank$.}
\begin{align}
    p_{c,B}(z)=&\;\P\bigg(\frac{\xi_B\,G_B^*\,\mathcal{D}_{BB}^{-\alpha_B}(z)}{J}>\tau\bigg)
    =\P\bigg(G_B^*>\frac{\tau\,J}{\xi_B\,\mathcal{D}_{BB}^{-\alpha_B}(z)}\bigg),
\end{align}
with $J=\sigma_n^2+I$.
By definition, the CCDF of the Gamma distribution is $\Bar{F}_G(g)=\frac{\Gamma^u(m,\,m\,g)}{\Gamma(m)}\blank$, where $\Gamma^u(m,m\,g)=\integral_{m\,g}^\infty t^{m-1}\,e^{-t}\,{\rm d}t$ is the upper incomplete Gamma function.
Let $\mu_B(z)=m_B\,\frac{\tau}{\xi_B}\,\mathcal{D}_{BB}^{\alpha_B}(z)$, taking the expectation with respect to $J$ implies that \cite{Alzenad19}
\begin{align}
    p_{c,B}(z)=&\; \E_J\bigg[\frac{\Gamma^u\big(m_B,\mu_B(z)\,J\big)}{\Gamma(m_B)}\bigg] 
    \overset{(a)}{=} \E_J\bigg[e^{-\mu_B(z)\,J}\,\sum\limits_{k=0}^{m_B-1}\frac{\big(\mu_B(z)\,J\big)^k}{k!}\bigg] \nonumber\\
    \overset{(b)}{=}&\; \sum\limits_{k=0}^{m_B-1}\frac{\mu_B^k(z)}{k!}\,\E_J\left[e^{-\mu_B(z)\,J}\,J^k\right],
\end{align}
where $(a)$ follows from the definition $\frac{\Gamma^u(m,g)}{\Gamma(m)}=e^{-g}\sum\limits_{k=0}^{m-1}\frac{g^k}{k!}$, and $(b)$ is obtained from the linearity of the expectation operator.
Taking into account that 
\begin{align}
\E_J\big[e^{-s\,J}\,J^k\big]=(-1)^k\,\frac{\partial^k}{\partial s^k}\,\mathcal{L}_J(s|z), \nonumber
\end{align}
where 
\begin{align}\mathcal{L}_{J}(s|z)=&\; \E\big[e^{-s\,J}\big]=\E\big[e^{-s\,I}\,e^{-s\sigma_n^2}\big] 
= e^{-s\,\sigma_n^2}\,\E\big[e^{-s\,I}\big]=e^{-s\,\sigma_n^2}\,\mathcal{L}_{I}(s|z)\blank,\nonumber
\end{align} the final expression is obtained. 
This, however, may require the computation of high-order derivatives of the conditional Laplace transform of the interference, resulting in a number of terms proportional to $m_B\blank$.

\section{Proof of Theorem~\ref{thm:Pc_tilde}} \label{appx:Pc_tilde}
A tight bound can be applied to the CDF of the Gamma distribution in order to ease the computation of the conditional coverage probabilities provided in Theorem~\ref{thm:Pc}. 
Let 
$\Gamma_l(m,m\,g)=\integral_0^{m\,g}t^{m-1}\,e^{-t}\,{\rm d}t$
denote the lower incomplete Gamma function, then the CDF of the Gamma distribution $F_{G}(g)=\frac{\Gamma_l(m,\,m\,g)}{\Gamma(m)}=1-\frac{\Gamma^u(m,\,m\,g)}{\Gamma(m)}$, 
can be bounded as \cite{alzer97some}
$$(1-e^{-\varepsilon_{1}\,m\,g})^{m}\leq\frac{\Gamma_l(m,m\,g)}{\Gamma(m)}\leq(1-e^{-\varepsilon_{2}\,m\,g})^{m},$$
where we defined the constants $\varepsilon_{1}=\begin{cases}
 1, & \text{if } m\geq1 \nonumber\\ (m!)^{-\frac{1}{m}}, & \text{if } m<1   \end{cases}$ and $\varepsilon_{2}=\begin{cases}
 (m!)^{-\frac{1}{m}}, & \text{if } m>1 \nonumber\\ 1, & \text{if } m\leq1   \end{cases}\,$. \par
Note that for $m=1$ the upper and the lower bounds become equal and thus $\frac{\Gamma_l(1,g)}{\Gamma(1)}=1-e^{-g}$.
It has been shown in \cite{bai15coverage} that the upper bound actually is a good approximation, hence we consider $\varepsilon_{2}=(m!)^{-\frac{1}{m}}$. \par
Recalling that $\mu_B(z)=m_B\,\frac{\tau}{\xi_B}\,\mathcal{D}_{BB}^{\alpha_B}(z)$, the conditional coverage probabilities can be approximated as \cite[Appendix F]{Alzenad19}
\begin{align}
    p_{c,B}=&\;\E_{J}\bigg[\frac{\Gamma^u(m_B,\mu_B(z)\,J)^k}{\Gamma(m_B)}\bigg] 
    = \E_{J}\bigg[1-\frac{\Gamma_l(m_B,\mu_B(z)\,J)}{\Gamma(m_B)}\bigg] 
    \overset{(a)}{\approx} 1-\E_{J}\Big[\left(1-e^{-\varepsilon_{2,B}\,\mu_B(z)\,J}\right)^{m_B}\Big] \nonumber\\
    \overset{(b)}{=}&\; 1-\E_{J}\bigg[\sum\limits_{k=0}^{m_B}\binom{m_B}{k}\,(-1)^{m_B-k}\,(-e^{-\varepsilon_{2,B}\,\mu_B(z)\,J})^k\bigg] \nonumber\\
    =&\; \E_{J}\bigg[\sum\limits_{k=1}^{m_B}\binom{m_B}{k}\,(-1)^{k+1}\,\exp(-k\,\varepsilon_{2,B}\,\mu_B(z)\,J)\bigg] \nonumber\\ 
    =&\;\sum\limits_{k=1}^{m_B}\binom{m_B}{k}\,(-1)^{k+1}\,\E_{J}\big[\exp\big(-k\,\varepsilon_{2,B}\,\mu_B(z)\,J\big)\big], 
\end{align}
where $(a)$ follows from the upper bound previously introduced and $(b)$ from the binomial theorem under the assumption that $m_B\!\in\!\mathbb{N}\blank$.
The final result in (\ref{eq:approx_PcB}) can be obtained by applying the definition of the conditional Laplace transform of the interference.

\bibliographystyle{IEEEtran}
\bibliography{Text.bib}

\begin{thebibliography}{10}
\providecommand{\url}[1]{#1}
\csname url@samestyle\endcsname
\providecommand{\newblock}{\relax}
\providecommand{\bibinfo}[2]{#2}
\providecommand{\BIBentrySTDinterwordspacing}{\spaceskip=0pt\relax}
\providecommand{\BIBentryALTinterwordstretchfactor}{4}
\providecommand{\BIBentryALTinterwordspacing}{\spaceskip=\fontdimen2\font plus
\BIBentryALTinterwordstretchfactor\fontdimen3\font minus
  \fontdimen4\font\relax}
\providecommand{\BIBforeignlanguage}[2]{{%
\expandafter\ifx\csname l@#1\endcsname\relax
\typeout{** WARNING: IEEEtran.bst: No hyphenation pattern has been}%
\typeout{** loaded for the language `#1'. Using the pattern for}%
\typeout{** the default language instead.}%
\else
\language=\csname l@#1\endcsname
\fi
#2}}
\providecommand{\BIBdecl}{\relax}
\BIBdecl

\bibitem{INSARAG}
{United Nations Office for Coordination of Humanitarian Affairs (UNOCHA)},
  ``{The Story of INSARAG 20 Years On...}'' 2010.

\bibitem{esposito20reinforced}
C.~Esposito, Z.~Zhao, and J.~Rak, ``Reinforced secure gossiping against {DoS}
  attacks in post-disaster scenarios,'' \emph{IEEE Access}, vol.~8, pp.
  178\,651--178\,669, 2020.

\bibitem{matracia22survey}
M.~Matracia, N.~Saeed, M.~A. Kishk, and M.-S. Alouini, ``Post-disaster
  communications: Enabling technologies, architectures, and open challenges,''
  \emph{IEEE Open Journal of the Communications Society}, vol.~3, pp.
  1177--1205, 2022.

\bibitem{Matracia21disaster}
M.~Matracia, M.~A. Kishk, and M.-S. Alouini, ``On the topological aspects of
  {UAV}-assisted post-disaster wireless communication networks,'' \emph{IEEE
  Communications Magazine}, vol.~59, no.~11, pp. 59--64, 2021.

\bibitem{Mozaffari19tutorial}
M.~{Mozaffari}, W.~{Saad}, M.~{Bennis}, Y.~{Nam}, and M.~{Debbah}, ``A tutorial
  on {UAV}s for wireless networks: Applications, challenges, and open
  problems,'' \emph{IEEE Communications Surveys Tutorials}, vol.~21, no.~3, pp.
  2334--2360, 2019.

\bibitem{al2014optimal}
A.~Al-Hourani, S.~Kandeepan, and S.~Lardner, ``Optimal {LAP} altitude for
  maximum coverage,'' \emph{IEEE Wireless Communications Letters}, vol.~3,
  no.~6, pp. 569--572, 2014.

\bibitem{savkin20navigation}
A.~V. Savkin and H.~Huang, ``Navigation of a network of aerial drones for
  monitoring a frontier of a moving environmental disaster area,'' \emph{IEEE
  Systems Journal}, vol.~14, no.~4, pp. 4746--4749, 2020.

\bibitem{wu18coupling}
W.~Wu, M.~A. Qurishee, J.~Owino, I.~Fomunung, M.~Onyango, and B.~Atolagbe,
  ``Coupling deep learning and {UAV} for infrastructure condition assessment
  automation,'' in \emph{IEEE International Smart Cities Conference (ISC2)},
  Kansas City, Missouri, USA, 2018, pp. 1--7.

\bibitem{sanjana20aid}
P.~Sanjana and M.~Prathilothamai, ``Drone design for first aid kit delivery in
  emergency situation,'' in \emph{6th International Conference on Advanced
  Computing and Communication Systems (ICACCS)}, Piscataway, New Jersey, USA,
  2020, pp. 215--220.

\bibitem{Kishk20magazine}
M.~{Kishk}, A.~{Bader}, and M.-S. {Alouini}, ``Aerial base station deployment
  in {6G} cellular networks using tethered drones: The mobility and endurance
  tradeoff,'' \emph{IEEE Vehicular Technology Magazine}, vol.~15, no.~4, pp.
  103--111, 2020.

\bibitem{kishk20placement}
M.~A. {Kishk}, A.~{Bader}, and M.-S. {Alouini}, ``On the {3-D} placement of
  airborne base stations using tethered {UAVs},'' \emph{IEEE Transactions on
  Communications}, vol.~68, no.~8, pp. 5202--5215, 2020.

\bibitem{Erdelj16_disasterManagement}
M.~Erdelj and E.~Natalizio, ``{UAV}-assisted disaster management: Applications
  and open issues,'' in \emph{International Conference on Computing, Networking
  and Communications (ICNC)}.\hskip 1em plus 0.5em minus 0.4em\relax IEEE
  Computer Society, Kauai, Hawaii, USA, 2016, pp. 1--5.

\bibitem{Graven17managing}
O.~H. Graven, J.-V. S{\o}rli, J.~Bj{\o}rk, D.~A.~H. Samuelsen, and J.~D.
  Bjerknes, ``Managing disasters-rapid deployment of sensor network from
  drones: Providing first responders with vital information,'' in \emph{2nd
  International Conference on Control and Robotics Engineering (ICCRE)}.\hskip
  1em plus 0.5em minus 0.4em\relax IEEE, Bangkok, Thailand, 2017, pp. 184--188.

\bibitem{ezequiel2014uav}
C.~A.~F. Ezequiel, M.~Cua, N.~C. Libatique, G.~L. Tangonan, R.~Alampay, R.~T.
  Labuguen, C.~M. Favila, J.~L.~E. Honrado, V.~Canos, C.~Devaney, A.~Loreto,
  J.~Bacusmo, and B.~Palma, ``{UAV aerial imaging applications for
  post-disaster assessment, environmental management and infrastructure
  development},'' in \emph{International Conference on Unmanned Aircraft
  Systems (ICUAS)}.\hskip 1em plus 0.5em minus 0.4em\relax IEEE, Orlando,
  Florida, USA, 2014, pp. 274--283.

\bibitem{Naqvi18key}
S.~A.~R. Naqvi, S.~A. Hassan, H.~Pervaiz, and Q.~Ni, ``Drone-aided
  communication as a key enabler for {5G} and resilient public safety
  networks,'' \emph{IEEE Communications Magazine}, vol.~56, no.~1, pp. 36--42,
  2018.

\bibitem{Selim18battery}
M.~Y. Selim and A.~E. Kamal, ``Post-disaster {4G/5G} network rehabilitation
  using drones: Solving battery and backhaul issues,'' in \emph{IEEE Globecom
  Workshops (GC Wkshps)}, Abu Dhabi, UAE, 2018, pp. 1--6.

\bibitem{shakoor19role}
S.~Shakoor, Z.~Kaleem, M.~I. Baig, O.~Chughtai, T.~Q. Duong, and L.~D. Nguyen,
  ``Role of {UAVs} in public safety communications: Energy efficiency
  perspective,'' \emph{IEEE Access}, vol.~7, pp. 140\,665--140\,679, 2019.

\bibitem{masroor21efficient}
R.~Masroor, M.~Naeem, and W.~Ejaz, ``Efficient deployment of {UAVs} for
  disaster management: A multi-criterion optimization approach,''
  \emph{Computer Communications}, vol. 177, pp. 185--194, 2021.

\bibitem{arshad2018integrating}
R.~Arshad, L.~Lampe, H.~ElSawy, and M.~J. Hossain, ``Integrating {UAVs} into
  existing wireless networks: A stochastic geometry approach,'' in \emph{IEEE
  Globecom Workshops (GC Wkshps)}, Abu Dhabi, UAE, 2018, pp. 1--6.

\bibitem{Alzenad19}
M.~{Alzenad} and H.~{Yanikomeroglu}, ``Coverage and rate analysis for vertical
  heterogeneous networks ({VHetNets}),'' \emph{IEEE Transactions on Wireless
  Communications}, vol.~18, no.~12, pp. 5643--5657, 2019.

\bibitem{Matracia21rural}
M.~Matracia, M.~A. Kishk, and M.-S. Alouini, ``Coverage analysis for
  {UAV}-assisted cellular networks in rural areas,'' \emph{IEEE Open Journal of
  Vehicular Technology}, vol.~2, pp. 194--206, 2021.

\bibitem{Kouzayha20hybrid}
N.~Kouzayha, H.~ElSawy, H.~Dahrouj, K.~Alshaikh, T.~Y. Al-Naffouri, and M.-S.
  Alouini, ``Stochastic geometry analysis of hybrid aerial terrestrial networks
  with mm{W}ave backhauling,'' in \emph{IEEE International Conference on
  Communications (ICC)}, Dublin, Ireland, 2020, pp. 1--7.

\bibitem{hayajneh18performance}
A.~M. Hayajneh, S.~A.~R. Zaidi, D.~C. McLernon, M.~Di~Renzo, and M.~Ghogho,
  ``Performance analysis of {UAV} enabled disaster recovery networks: A
  stochastic geometric framework based on cluster processes,'' \emph{IEEE
  Access}, vol.~6, pp. 26\,215--26\,230, 2018.

\bibitem{afshang17bpp}
M.~{Afshang} and H.~S. {Dhillon}, ``Fundamentals of modeling finite wireless
  networks using binomial point process,'' \emph{IEEE Transactions on Wireless
  Communications}, vol.~16, no.~5, pp. 3355--3370, 2017.

\bibitem{Chetlur3D_bpp}
V.~V. Chetlur and H.~S. Dhillon, ``Downlink coverage analysis for a finite
  {3-D} wireless network of unmanned aerial vehicles,'' \emph{IEEE Transactions
  on Communications}, vol.~65, no.~10, pp. 4543--4558, 2017.

\bibitem{Andrews11}
J.~G. {Andrews}, F.~{Baccelli}, and R.~K. {Ganti}, ``A tractable approach to
  coverage and rate in cellular networks,'' \emph{IEEE Transactions on
  Communications}, vol.~59, no.~11, pp. 3122--3134, 2011.

\bibitem{haenggi12stochastic}
M.~Haenggi, \emph{Stochastic Geometry for Wireless Networks}.\hskip 1em plus
  0.5em minus 0.4em\relax Cambridge University Press, 2012.

\bibitem{PRIMER}
\BIBentryALTinterwordspacing
J.~G. Andrews, A.~K. Gupta, and H.~S. Dhillon, ``A primer on cellular network
  analysis using stochastic geometry,'' 2016. [Online]. Available:
  \url{https://arxiv.org/abs/1604.03183}
\BIBentrySTDinterwordspacing

\bibitem{Galkin19}
B.~{Galkin}, J.~{Kibilda}, and L.~A. {DaSilva}, ``A stochastic model for {UAV}
  networks positioned above demand hotspots in urban environments,'' \emph{IEEE
  Transactions on Vehicular Technology}, vol.~68, no.~7, pp. 6985--6996, 2019.

\bibitem{alzer97some}
H.~Alzer, ``On some inequalities for the incomplete gamma function,''
  \emph{Mathematics of Computation}, vol.~66, no. 218, pp. 771--778, 1997.

\bibitem{bai15coverage}
T.~Bai and R.~W. Heath, ``Coverage and rate analysis for millimeter-wave
  cellular networks,'' \emph{IEEE Transactions on Wireless Communications},
  vol.~14, no.~2, pp. 1100--1114, 2015.

\end{thebibliography}
%
\begin{IEEEbiographynophoto}
{Maurilio Matracia}
[S'21] received his M.Sc. degree in Electrical Engineering from the University of Palermo (UNIPA), Italy, in 2019.
He is currently a Doctoral Student at the Communication Theory Lab (CTL), King Abdullah University of Science and Technology (KAUST), Kingdom of Saudi Arabia (KSA).
His main research interest is stochastic geometry, with a special focus on rural and emergency communications.
\end{IEEEbiographynophoto}


\begin{IEEEbiographynophoto}
{Mustafa A. Kishk}
[S'16, M'18] received his Ph.D. degree in Electrical Engineering from Virginia Tech, USA, in 2018.
He is currently an Assistant Professor with the Electronic Engineering Department, Maynooth University, Ireland.
His research interests include stochastic geometry, energy harvesting wireless networks, UAV-enabled communication systems, and satellite communications.
His current research interests include stochastic geometry, energy-harvesting wireless networks, UAV-enabled communication systems, and satellite communications.
\end{IEEEbiographynophoto}


\begin{IEEEbiographynophoto}
{Mohamed-Slim Alouini} 
[S'94, M'98, SM'03, F'09] was born in Tunis, Tunisia.
He received his Ph.D. degree in Electrical Engineering from California
Institute of Technology (Caltech), Pasadena, CA, USA. 
He served as a faculty member at the University of Minnesota, Minneapolis, MN, USA, then at Texas A$\&$M University at Qatar, Education City, Doha, Qatar, before joining KAUST as a Professor of Electrical Engineering in 2009. 
His current research interests include modeling, design, and performance analysis of wireless communication systems.
\end{IEEEbiographynophoto}

\end{document}